\documentclass[aps,prb,twocolumn,floatfix,superscriptaddress,longbibliography]{revtex4-2}
\usepackage{mathrsfs}
\usepackage{graphicx}
\usepackage[english]{babel}
\usepackage{amsmath}
\usepackage{amssymb}
\usepackage{xcolor}
\usepackage{relsize}
\usepackage{dsfont}
\usepackage{bm}% bold math
\graphicspath{{./}}
\usepackage{epstopdf}

\newcommand{\me}{\mathrm{e}}
\newcommand{\mi}{\mathrm{i}}

\newcommand{\dif}{\mathrm{d}}

\allowdisplaybreaks
\addto\captionsenglish{}
\begin{document}
	\title{Geometric and topological obstructions to global proper Hermitianization in quasi-Hermitian quantum systems}
	\author{Ming-Zhang Wang}
	\affiliation{School of Physics, Southeast University, Jiulonghu Campus, Nanjing 211189, China}
	\author{Xu-Yang Hou}
	\affiliation{School of Physics, Southeast University, Jiulonghu Campus, Nanjing 211189, China}
	\author{Hao Guo}
	\email{guohao.ph@seu.edu.cn}
	\affiliation{School of Physics, Southeast University, Jiulonghu Campus, Nanjing 211189, China}
	\affiliation{Hefei National Laboratory, University of Science and Technology of China, Hefei 230088, China}
	\begin{abstract}
		
		Quasi-Hermitian quantum systems with a specified positive-definite physical metric and metric-compatible dynamics can be mapped instantaneously to equivalent Hermitian systems via a similarity transformation.
		Extending this equivalence consistently over a multidimensional parameter space requires compatibility among the proper similarity transformations obtained along different paths. We show that this compatibility can be characterized by a metric-induced connection. The curvature of this connection characterizes the local geometric obstruction to extending pathwise proper transformations to a common proper frame. Its holonomy, in turn, characterizes the global topological obstruction to single-valuedness that may occur even when the connection is flat. Finally, through four two-level models, we illustrate how the geometric and topological obstructions can be diagnosed in concrete systems and demonstrate observable endpoint signatures arising from nontrivial holonomy together with possible experimental probes.
	\end{abstract}
	\maketitle
	\section{Introduction}

	Non-Hermitian Hamiltonians appear in several distinct physical settings,
	including effective descriptions of open, dissipative, gain--loss, and
	post-selected systems, as well as quasi-Hermitian theories formulated
	with a nonstandard physical inner product
	~\cite{FESHBACH1958357,Moiseyev_book,El_Ganainy_2018}.
	
	Their study has revealed unconventional spectral and dynamical behavior
	as well as rich topological phenomena, including the non-Hermitian skin
	effect and modified bulk--boundary correspondences
	~\cite{PhysRevX.8.031079,PhysRevLett.121.026808,Roccati2023,
		PhysRevLett.121.086803,Li2019NatCommun,PhysRevLett.123.230401}.
	These developments have stimulated extensive interest in the geometric
	and topological structures of non-Hermitian quantum systems.
	
	$\mathcal{PT}$-symmetric Hamiltonians form a widely studied class
	of non-Hermitian models. In the unbroken regime, a diagonalizable
	$\mathcal{PT}$-symmetric Hamiltonian with a real spectrum may admit a
	positive-definite metric and an equivalent Hermitian representation
	~\cite{PhysRevLett.110.083604,RevModPhys.88.035002,
		PhysRevB.82.052404,Bender24,Cham_2015,Feng_2017,Korff_2007,Korff_2008,Bender1999,PhysRevA.92.032106}.
	This metric defines an inner product in which the Hamiltonian is self-adjoint, providing the quasi-Hermitian framework
	~\cite{RevModPhys.15.175,PhysRevLett.80.5243,
		Mostafazadeh2002I,Mostafazadeh2002II,Mostafazadeh2002III}.
	Here the positive metric is taken to define the physical inner product, and state evolution follows the corresponding metric-compatible dynamics.
	
	Recent digital quantum simulation provides access to time-dependent metric formalisms through dilation or embedding~\cite{Gonzalez2026Emulation}. Such emulation offers a route to studying a specified quasi-Hermitian structure \((H,\eta)\); it is distinct from a native experimental realization in which the metric itself defines the physical inner product.
	
	Because the physical inner product $(\cdot,\cdot)_\eta=\langle\cdot|\eta|\cdot\rangle$ may depend on external parameters, the geometry of the physical Hilbert space is itself parameter dependent~\cite{doi:10.1142/S0219887810004816,Das_2011}. This has led to extensions of geometric phases to quasi-Hermitian and related non-Hermitian systems~\cite{JPAGW13,PhysRevB.109.245411}, as well as to a formulation of the corresponding quantum geometric tensor~\cite{PhysRevA.99.042104}. In what follows, the metric $\eta(\boldsymbol{R})$ is taken as part of the system's physical specification, so the obstruction criteria refer to the given quasi-Hermitian structure $(H,\eta)$ rather than to $H$ alone.
	
	A natural and powerful strategy to analyze quasi-Hermitian systems is to map them, via a similarity transformation $S$, to an equivalent Hermitian system. Two conceptually distinct frameworks for Hermitianization exist. The first is an instantaneous algebraic procedure: for each fixed parameter value, one chooses $S$ such that $\eta=S^\dagger S$ and $\tilde{H}=SHS^{-1}$ is Hermitian. A simple example is $S=\sqrt{\eta}$. The second framework, which is the main subject of this work, imposes the additional requirement that $S_p$ convert the modified quasi-Hermitian Schr\"odinger equation into the ordinary Schr\"odinger equation for the Hermitian system. This dynamical compatibility is encoded in the proper condition $\dot S_p^\dagger S_p=S_p^\dagger\dot S_p$~\cite{JPAGW13}.
	
	The construction builds on several established results. The metric-based pseudo-Hermitian formalism and the relation between positive metrics and equivalent Hermitian representations were developed systematically in Refs.~\cite{Mostafazadeh2002I,Mostafazadeh2002II,Mostafazadeh2002III,Mostafazadeh2008Metric}. Time-dependent quasi-Hermitian and $\mathcal{PT}$-symmetric dynamics, proper similarity transformations, and the mapping of a closed quasi-Hermitian parameter path to an open Hermitian one were studied in Ref.~\cite{JPAGW13}. The corresponding quantum geometric tensor was formulated in Ref.~\cite{PhysRevA.99.042104}, while time-dependent Dyson maps and metric-dependent dynamics were examined in Ref.~\cite{PhysRevA.94.042128}.		
	Once a proper transformation $S_p$ is available, the physical properties
	of the quasi-Hermitian system, including spectra, dynamics, and geometric
	phases, can be related to those of an equivalent Hermitian
	system~\cite{PhysRevA.99.042104,PhysRevA.94.042128}.
	
	Here we focus on a different issue: the compatibility of proper transformations obtained along different paths, specifically whether they can be combined into a common, globally single-valued proper frame over a multidimensional parameter domain. We refer to the failure of such a combination as an obstruction to global proper Hermitianization. This notion leaves intact both instantaneous algebraic Hermitianization and the existence of a proper transformation along any prescribed one-dimensional path.
	
	The metric $\eta$ induces a natural gauge connection $\mathcal{G}_\mu = \frac{1}{2}[\partial_\mu\sqrt{\eta},\sqrt{\eta}^{-1}]$. Its curvature characterizes the local compatibility of proper transformations across different directions, whereas its holonomy characterizes their global consistency around closed loops. Nonzero curvature prevents these from extending to a common proper frame over a neighborhood; even vanishing curvature can permit nontrivial holonomy around a noncontractible loop, obstructing global single-valuedness. The closed-to-open behavior found in Ref.~\cite{JPAGW13} is then an endpoint holonomy of this connection.
	
	These implications are particularly clear along a closed parameter path. Although the parameters return to their initial values, the pathwise Hermitian representation need not; its endpoint may differ from the initial one by a sewing unitary determined by the Wilson loop. The mapped Hermitian Hamiltonian is periodic when the sewing unitary commutes with the initial Hermitian Hamiltonian.
	The resulting rotation of the Hamiltonian and its eigenstates preserves the spectrum while changing their orientation relative to a fixed measurement axis. Adiabatic endpoint measurements reveal this rotation when the initial proper frame and reference observable are held fixed.
	
	To illustrate these mechanisms, we first consider three examples: (i) an integrable quasi-Hermitian structure with a globally single-valued proper frame; (ii) a disk with nonzero curvature, where a proper transformation exists along every circular trajectory but generally fails to return to its initial value after one cycle; and (iii) an annular domain with vanishing curvature but a nontrivial Wilson loop around the hole. In the third and fourth examples, distinct closed paths with the same initial and final parameters lead to different holonomies and hence different endpoint readouts in a common Hermitian reference frame. These examples also suggest possible experimental probes of the corresponding geometric effects. The fourth example further demonstrates that, after calibrated endpoint alignment, the two paths yield a measurable relative phase in their Hermitian images.
	Our results establish geometric and topological criteria for the existence of a global proper Hermitian frame for a fixed quasi-Hermitian structure.
	
	The paper is organized as follows. Section~\ref{Sec.1} introduces the quasi-Hermitian structure, distinguishes the instantaneous energy operator from the dynamical generator, and reviews the associated Berry geometry. Section~\ref{Sec.2} discusses the proper similarity transformation, its existence along a prescribed trajectory, and its uniqueness up to a unitary factor that is constant along the trajectory. Section~\ref{Sec.3} formulates the multidirectional extension problem, introduces a directional curvature mismatch, and derives the operator-level integrability condition. Section~\ref{Sec.4} treats the global holonomy obstruction on non-simply-connected parameter domains. The four examples are presented in Sec.~\ref{Sec.5}, followed by the conclusion in Sec.~\ref{Sec.6}. Technical details are collected in the Appendixes.

	\section{Framework of quasi-Hermitian quantum mechanics}\label{Sec.1}
	
	\subsection{Quasi-Hermitian framework and Berry geometry}
	We consider a finite-dimensional parameter-dependent quasi-Hermitian structure $(H(\boldsymbol R),\eta(\boldsymbol R))$, with $\boldsymbol R=(R^1,R^2,\ldots)$. The operator $H(\boldsymbol R)$ plays the role of the instantaneous energy operator. Although it is generally non-Hermitian with respect to the standard Dirac inner product, it is self-adjoint with respect to the physical inner product defined by $\eta$ when
	\begin{equation}\label{eq:formula_001}
		\eta(\boldsymbol{R}) H(\boldsymbol{R}) = H^{\dagger}(\boldsymbol{R}) \eta(\boldsymbol{R}) .
	\end{equation}
	A smooth positive-definite metric
	$\eta(\boldsymbol R)$ is assumed on the parameter manifold.
	The quasi-Hermiticity relation then implies that
	$H(\boldsymbol R)$ has a real spectrum and is diagonalizable, so that a
	complete biorthonormal set of right and left eigenvectors can be chosen.
	Units with $\hbar=1$ are used throughout. Let $|\Psi_n(\boldsymbol{R})\rangle$ and $|\Phi_n(\boldsymbol{R})\rangle$ denote the right and left eigenvectors respectively,
	\begin{align}\label{eq:formula_002}
		H(\boldsymbol{R}) |\Psi_n(\boldsymbol{R})\rangle &= E_n(\boldsymbol{R}) |\Psi_n(\boldsymbol{R})\rangle ,\notag\\
		H^{\dagger}(\boldsymbol{R}) |\Phi_n(\boldsymbol{R})\rangle &= E_n(\boldsymbol{R}) |\Phi_n(\boldsymbol{R})\rangle ,
	\end{align}
	with the biorthonormality condition $\langle\Phi_n(\boldsymbol{R})|\Psi_m(\boldsymbol{R})\rangle = \delta_{nm}$. The metric $\eta$ connects the two bases via
	\begin{equation}\label{eq:formula_003}
		\eta(\boldsymbol{R}) |\Psi_n(\boldsymbol{R})\rangle = |\Phi_n(\boldsymbol{R})\rangle .
	\end{equation}
	The physical inner product is defined as $(\cdot,\cdot)_\eta = \langle\cdot|\eta|\cdot\rangle$. This inner product is positive-definite and, when combined with the appropriate time evolution equation, guarantees unitarity and a consistent probabilistic interpretation.
	
	The metric is part of the physical definition of the system. In particular, it determines the observable algebra
	
	\begin{equation}\label{observable_algebra}
		\mathfrak A _\eta
		=
		\left\{O\,\middle|\,O^\dagger\eta=\eta O\right\}.
	\end{equation}
	
	Accordingly, two positive metrics $\eta_1$ and $\eta_2$
	satisfying $H^\dagger\eta_j=\eta_jH$ $(j=1, 2)$ for the same instantaneous energy
	operator $H$ generally define different inner products and
	observable algebras. They describe equivalent theories only when
	the corresponding full structures are related by an appropriate
	unitary isomorphism; the specification of $H$ alone is not
	sufficient~\cite{Scholtz1992,Geyer2007Metric,Mostafazadeh2008Metric}.
	
	For a time-dependent metric, however, the instantaneous
	energy operator $H$ is not by itself the generator of state evolution.
	Along a prescribed trajectory $\boldsymbol R(t)$, one has
	$\dot\eta=\dot R^\mu\partial_\mu\eta$. If one nevertheless used
	$\mi\partial_t|\Psi\rangle=H|\Psi\rangle$, taking the time derivative of
	the physical norm gives
	\begin{equation}\label{eq:formula_004}
		\frac{\dif}{\dif t}\langle\Psi|\eta|\Psi\rangle
		=
		\langle\Psi|\dot\eta|\Psi\rangle ,
	\end{equation}
	which is generally nonzero when the metric varies in time.
	We therefore take the state evolution to be governed by the metric-compatible Schr\"odinger equation~\cite{JPAGW13,Mostafazadeh2020TD}
	\begin{equation}\label{5}
		\mi\frac{\dif}{\dif t}|\Psi\rangle = \left( H - \frac{\mi}{2}\eta^{-1}\dot\eta \right) |\Psi\rangle .
	\end{equation}
	The operator in parentheses generates the state evolution,
	whereas $H$ determines the instantaneous energy spectrum. This
	metric-compatible evolution conserves the physical norm under the
	quasi-Hermiticity condition. The proper transformations introduced below are defined with respect to this dynamics.

	Norm conservation holds for each admissible metric together with its corresponding generator, but does not single out a unique metric. For two admissible choices,

	\begin{equation}\label{metric_dependent_generators}
		K_{\eta_j}
		=
		H-\frac{\mi}{2}\eta_j^{-1}\dot\eta_j,
		\qquad j=1,2,
	\end{equation}
	
	which are generally different. Here $K_{\eta_j}$ denotes the full dynamical generator and is unrelated to the parameter-space quantity $\mathcal K_\mu$ introduced below.
	The pair \((H,\eta)\) thus specifies both the physical Hilbert space and the metric-compatible evolution~\cite{Mostafazadeh2020TD}. Equation~\eqref{5} provides the dynamics for the geometric construction below.
	In effective descriptions of open, gain--loss, dissipative, or post-selected systems, the generator and physical inner product need not have this relation. Applying the construction in those settings requires a specified positive metric and the corresponding compatible dynamics.
	
	For an adiabatic change of parameters, the geometric phase acquired by a non-degenerate eigenstate $|\Psi_n(\boldsymbol{R})\rangle$ was derived in Ref.~\cite{PhysRevA.99.042104}. This phase is used to compare the pathwise Hermitian representations. The associated quasi-Hermitian Berry connection $\mathcal{A}_{n,\mu}(\boldsymbol{R})$ is given by
	\begin{align}\label{BerryA}
		\mathcal{A}_{n,\mu}(\boldsymbol{R}) =& \mi\Big( \langle\Psi_n(\boldsymbol{R})|\eta(\boldsymbol{R}) \partial_\mu |\Psi_n(\boldsymbol{R})\rangle\notag\\
		+& \frac12 \langle\Psi_n(\boldsymbol{R})| \partial_\mu\eta(\boldsymbol{R}) |\Psi_n(\boldsymbol{R})\rangle \Big) ,
	\end{align}
	where $\partial_\mu \equiv \partial/\partial R^\mu$. This connection is real and transforms as a gauge connection under a phase redefinition of the instantaneous eigenstate.
	For a single-valued periodic eigenstate, the corresponding Berry phase is
	$\gamma_n=\mathlarger{\oint}\mathcal A_{n,\mu}\dif R^\mu$, which is
	gauge invariant modulo $2\pi$. The presence of the metric $\eta$ and its
	derivative in the connection reflects the parameter dependence of the
	physical inner product.
	If the eigenvector chosen continuously along the
	loop does not return to its initial value after a cyclic parameter
	evolution, the gauge-invariant geometric phase must also include an
	endpoint-overlap contribution.

	\subsection{Geometric preliminaries}

	For a nondegenerate eigenstate, the Berry connection can be viewed as a
	$U(1)$ connection on the corresponding eigenstate bundle over parameter
	space, with the Berry curvature encoding its local geometric
	structure~\cite{Simon1983}. For degenerate eigenspaces or, more generally,
	the transport of a multi-component frame, the connection becomes matrix
	valued and gives rise to a non-Abelian gauge structure
	~\cite{WilczekZee1984,Nakahara2003}. The associated gauge curvature
	characterizes the local noncommutativity of parallel transport in
	different parameter directions. In the present work, the Berry curvature associated with the instantaneous eigenstates should be distinguished from the curvature of the
	metric-induced matrix-valued connection introduced in Sec.~\ref{sec:metric_gauge_connection}: the
	former characterizes the geometry of individual eigenstates, whereas
	the latter governs the local compatibility of proper transformations in
	different parameter directions.

	A matrix-valued connection defines parallel transport along a path in
	parameter space~\cite{Nakahara2003}. The resulting path-ordered transport
	operator is called a Wilson line for an open path and a Wilson loop for
	a closed path. For a closed circuit, the Wilson loop represents the
	holonomy of the connection, namely the transformation relating the
	transported frame at the end of the circuit to its initial value. If
	this transformation is nontrivial, the transported frame fails to
	return to its original value after one circuit and exhibits a
	monodromy.

	Curvature and holonomy therefore characterize complementary local and
	global aspects of parallel transport. A flat connection on a simply
	connected smooth domain is pure gauge and has trivial holonomy, whereas
	on a non-simply-connected domain a flat connection may still possess
	nontrivial holonomy around a noncontractible loop
	~\cite{Nakahara2003}. For a contractible closed contour, the
	non-Abelian Stokes theorem relates the Wilson loop to the curvature
	enclosed by the contour~\cite{Fishbane1981,Chepelev2002}. This
	distinction between local curvature and global holonomy underlies the
	geometric and topological obstructions discussed below.

	\section{Proper similarity transformation}\label{Sec.2}
	
	We now consider a natural strategy to analyze a quasi-Hermitian system: seeking an invertible operator $S$ such that
	\begin{equation}\label{eq:formula_005}
		\eta = S^\dagger S , \qquad \tilde{H} = S H S^{-1}
	\end{equation}
	is Hermitian. The eigenvectors of $\tilde{H}$ are $|\psi_n\rangle = S|\Psi_n\rangle$, which satisfy the standard orthonormality $\langle\psi_n|\psi_m\rangle = \delta_{nm}$ because $\langle\Psi_n|\eta|\Psi_m\rangle = \delta_{nm}$. If such an $S$ can be found, the quasi-Hermitian system can be analyzed in parallel with the Hermitian system described by $\tilde{H}$. However, the simple substitution $|\psi_n\rangle = S|\Psi_n\rangle$ does not automatically guarantee that the transformed evolving state obeys the ordinary Schr\"odinger equation $\mi\partial_t|\psi_n\rangle = \tilde{H}|\psi_n\rangle$. The label $n$ is retained for later comparison with instantaneous eigenstates, but the following transformation applies to the full evolution generated by Eq.~\eqref{5}; the evolving state need not remain an instantaneous eigenstate.
	Instead, when the similarity transformation additionally satisfies the proper condition, we denote it by $S_p$. Using the modified Schr\"odinger equation \eqref{5} and $\eta=S_p^\dagger S_p$, together with the proper condition
	introduced in Ref.~\cite{JPAGW13},
	\begin{equation}\label{PofS}
		\dot S_p^\dagger S_p=S_p^\dagger\dot S_p ,
	\end{equation}
	we have
	$
	\eta^{-1}\dot\eta
	=
	S_p^{-1}(S_p^\dagger)^{-1}\dot S_p^\dagger S_p
	+
	S_p^{-1}\dot S_p
	=
	2S_p^{-1}\dot S_p .
	$
	Hence the evolution equation for $|\Psi_n\rangle$ becomes
	$
	\mi\frac{\dif}{\dif t}|\Psi_n\rangle
	=
	\left(H-\mi S_p^{-1}\dot S_p\right)|\Psi_n\rangle .
	$
	Multiplying both sides by $S_p$ gives
	\begin{equation}\label{psiH}
		\mi\frac{\dif}{\dif t}|\psi_n\rangle
		=
		S_pHS_p^{-1}|\psi_n\rangle
		=
		\tilde H|\psi_n\rangle ,
	\end{equation}
	which is the standard Schr\"odinger equation for the Hermitian system. All steps are reversible; therefore this proper condition is not merely a convenient gauge choice but the necessary and sufficient condition for the similarity transformation to convert the quasi-Hermitian dynamics into ordinary Hermitian dynamics. Note that this condition is not required for the instantaneous Hermiticity of $\tilde H$.
	
	The remaining question is whether a proper $S_p$ exists along a given trajectory for an arbitrary $\eta(t)$. Writing $S_p=U\sqrt{\eta}$ with $U$ unitary, the condition \eqref{PofS} is equivalent to a first-order differential equation for $U$,
	\begin{equation}\label{dU}
		\dot U = \frac12 U \bigl(  (\sqrt{\eta})^{-1}\dot{\sqrt{\eta}} -\dot{\sqrt{\eta}}\,(\sqrt{\eta})^{-1} \bigr) .
	\end{equation}
	The matrix multiplying $U$ on the right-hand side is anti-Hermitian, so unitarity is preserved. For any initial condition, such as $U(0)=I$, Eq.~\eqref{dU} has a unique solution along the trajectory, and hence a pathwise proper $S_p$ always exists. Moreover, as shown in Appendix~\ref{appdU}, any two proper forms for the same $\eta$ along the same trajectory differ by a unitary matrix $C$ that is constant on that trajectory (i.e., $S_{p,2} = C S_{p,1}$). Whether one transformation can be proper in several parameter directions is a separate integrability problem, considered in Sec.~\ref{Sec.3}.
	
	\section{Geometric obstruction}\label{Sec.3}
	
	\subsection{Directional properness and Berry-curvature mismatch}

	Since the proper condition originates from the
	time-dependent quasi-Hermitian Schr\"odinger equation, it is imposed
	along a path in parameter space. We construct $S_p$ by integrating the
	proper equation along a selected parameter direction $R^\mu$. With
	initial conditions chosen smoothly with respect to the remaining
	parameters, these solutions define a smooth family $S_p(\boldsymbol R)$, for which
	derivatives along the other parameter directions are well defined.
	The proper condition, however, is imposed only along $R^\mu$.
	Using Eq.~\eqref{PofS} together with $\eta=S_p^\dagger S_p$, we obtain

	\begin{equation}
		(\partial_\mu S_p^\dagger)S_p
		=
		S_p^\dagger\partial_\mu S_p
		=
		\frac{1}{2}\partial_\mu\eta .
		\label{eq:directional_proper}
	\end{equation}

	Setting $|\psi_n\rangle=S_p|\Psi_n\rangle$, the Hermitian-image Berry
	connection along the selected direction becomes

	\begin{align}
		\widetilde{\mathcal A}_{n,\mu}
		&=
		\mi\langle\psi_n|\partial_\mu\psi_n\rangle
		\nonumber\\
		&=
		\mi\left[
		\langle\Psi_n|\eta|\partial_\mu\Psi_n\rangle
		+
		\frac{1}{2}
		\langle\Psi_n|\partial_\mu\eta|\Psi_n\rangle
		\right]
		=
		\mathcal A_{n,\mu}.
		\label{eq:directional_connection_equality}
	\end{align}

	Thus, the quasi-Hermitian and Hermitian Berry connections coincide along
	the selected direction $\mu$, while their components along a transverse
	direction $\nu$ may differ. The corresponding properness
	mismatch is defined as

	\begin{equation}
		\mathcal M_\nu[S_p]
		\equiv
		S_p^\dagger\partial_\nu S_p
		-
		\frac{1}{2}\partial_\nu\eta .
		\label{eq:transverse_mismatch}
	\end{equation}

	This quantity measures the extent to which $S_p$, constructed
	along $\mu$, deviates from the proper condition along the transverse
	direction $\nu$.
	Accordingly, the Berry-connection difference is

	\begin{equation}
		\widetilde{\mathcal A}_{n,\nu}
		-
		\mathcal A_{n,\nu}
		=
		\mi
		\langle\Psi_n|
		\mathcal M_\nu[S_p]
		|\Psi_n\rangle .
		\label{eq:transverse_connection_difference}
	\end{equation}

	We next compare the corresponding Berry curvatures~\cite{Simon1983,WilczekZee1984}.
	For the quasi-Hermitian system, the Berry curvature is defined as
	$\mathcal F_{n,\mu\nu}
	=\partial_\mu\mathcal A_{n,\nu}
	-\partial_\nu\mathcal A_{n,\mu}$,
	with $\widetilde{\mathcal F}_{n,\mu\nu}$ defined analogously for the
	Hermitian image.
	Using $\widetilde{\mathcal A}_{n,\mu}=\mathcal A_{n,\mu}$ together with Eq.~\eqref{eq:transverse_connection_difference}, we obtain
	
	\begin{equation}
		\widetilde{\mathcal F}_{n,\mu\nu}
		-
		\mathcal F_{n,\mu\nu}
		=
		\mi\partial_\mu
		\langle\Psi_n|
		\mathcal M_\nu[S_p]
		|\Psi_n\rangle
		\equiv
		-2\Delta_{n,\mu\nu}.
		\label{eq:compact_curvature_difference}
	\end{equation}

	Here $\Delta_{n,\mu\nu}$ serves as a measure of the directional
	Berry-curvature mismatch. It is determined by the variation along
	$\mu$ of the transverse properness mismatch; its explicit expansion is
	given in Appendix~\ref{app:directional_mismatch}. If the same $S_p$ is also proper
	along $\nu$, then $\mathcal M_\nu[S_p]=0$, which implies

	\begin{equation}
		\Delta_{n,\mu\nu}=0,
		\qquad
		\widetilde{\mathcal F}_{n,\mu\nu}
		=
		\mathcal F_{n,\mu\nu}.
		\label{eq:vanishing_curvature_difference}
	\end{equation}

	Conversely, a nonzero $\Delta_{n,\mu\nu}$ shows that the smooth
	family $S_p$, while proper along $\mu$, cannot be extended to remain
	proper along $\nu$ as well. It therefore
	signals the incompatibility of this directional construction with
	multidirectional properness.

	The quantity $\Delta_{n,\mu\nu}$ depends on the chosen smooth family of
	pathwise proper transformations, including its transverse initial
	condition, and is therefore not an intrinsic property of $(H,\eta)$.
	Its vanishing for a single eigenstate may be accidental. A
	state-independent criterion for local integrability must instead be
	formulated in terms of the metric-induced connection, as discussed in
	the next subsection.

	\subsection{Gauge connection $\mathcal{G}_\mu$ and its integrability}\label{sec:metric_gauge_connection}

	The directional mismatch $\Delta_{n,\mu\nu}$ provides a state-dependent diagnostic. To obtain a state-independent criterion, we turn to the operator level. The component form of Eq.~\eqref{dU} reads
	\begin{equation}\label{dU2}
		\partial_\mu U = -U \mathcal{G}_\mu ,\quad
		\mathcal{G}_\mu = \frac12 [\partial_\mu\sqrt{\eta},\,\sqrt{\eta}^{-1}] ,
	\end{equation}
	where $\mathcal{G}_\mu$ is anti-Hermitian and therefore defines a gauge connection. Requiring the transformation to be proper in both the $\mu$ and $\nu$ directions means that Eq.~\eqref{dU2} must hold simultaneously for the two indices. The integrability condition is that the mixed partial derivatives commute: $\partial_{\mu}\partial_\nu U = \partial_{\nu}\partial_\mu U$. Using Eq.~\eqref{dU2}, this becomes
	\begin{equation}\label{eq:formula_010}
		\mathcal{F}_{\mu\nu}^{\mathcal{G}} \equiv \partial_\mu\mathcal{G}_\nu - \partial_\nu\mathcal{G}_\mu - [\mathcal{G}_\mu,\mathcal{G}_\nu] = 0 .
	\end{equation}
	When $\mathcal{F}_{\mu\nu}^{\mathcal{G}}\neq0$, the directional solutions cannot be combined into a single transformation satisfying the proper equations on a common neighborhood. This nonzero curvature constitutes the local geometric obstruction inherent to the metric of the quasi-Hermitian structure $(H,\eta)$.

	This local obstruction should not be confused with a failure of existence along individual paths. Equation~\eqref{dU} is an ordinary differential equation and therefore admits a unique solution for $U$ along any prescribed trajectory; the nonvanishing curvature does not affect pathwise solvability. Rather, the obstruction lies in the impossibility of combining these pathwise solutions into a single $U$ that satisfies the partial derivative relations simultaneously over a neighborhood of the parameter space.

	\subsection{Equivalent integrability conditions}
	
	The same extension problem can be formulated directly in terms of $S_p$ by requiring
	\begin{equation}\label{dS}
		\partial_\mu S_{p} = S_{p} \mathcal{K}_\mu ,
	\end{equation}
	together with the analogous equation in the $\nu$ direction, where $\mathcal{K}_\mu$ is a yet-undetermined operator. As shown in Appendix~\ref{app2}, the proper condition yields $\mathcal{K}_\mu = \frac12 \eta^{-1} \partial_\mu \eta$. Imposing commutativity of mixed partial derivatives leads to the integrability condition $\partial_{\mu}\partial_\nu S_{p} = \partial_{\nu}\partial_\mu S_{p}$. Substituting \eqref{dS} gives
	\begin{align}\label{eq:formula_011}
		0 &= [\partial_\mu,\partial_\nu]S_{p} = \partial_\mu(S_{p} \mathcal{K}_\nu) - \partial_\nu(S_{p} \mathcal{K}_\mu) \notag\\
		&= S_{p}\bigl( \partial_\mu \mathcal{K}_\nu - \partial_\nu \mathcal{K}_\mu + [\mathcal{K}_\mu, \mathcal{K}_\nu] \bigr) .
	\end{align}
	Thus the same $S_p$ extends locally to both directions if and only if
	\begin{equation}\label{21}
		\mathcal{F}^{\mathcal{K}}_{\mu\nu} \equiv \partial_\mu \mathcal{K}_\nu - \partial_\nu \mathcal{K}_\mu + [\mathcal{K}_\mu, \mathcal{K}_\nu] = 0 .
	\end{equation}
	Using $\mathcal{K}_\mu = \frac12 \eta^{-1}\partial_\mu\eta$, one can derive the identity (see Appendix~\ref{app2})
	\begin{equation}\label{22}
		\partial_\mu \mathcal{K}_\nu - \partial_\nu \mathcal{K}_\mu = -2[\mathcal{K}_\mu, \mathcal{K}_\nu] .
	\end{equation}
	Consequently, the condition $\mathcal{F}^{\mathcal{K}}_{\mu\nu}=0$ simplifies to $[\mathcal{K}_\mu, \mathcal{K}_\nu]=0$. In Appendix~\ref{app2} we prove the relation
	\begin{equation}\label{eq:formula_012}
		\mathcal{F}^{\mathcal{K}}_{\mu\nu} = -\,\sqrt{\eta}^{-1} \mathcal{F}^{\mathcal{G}}_{\mu\nu} \sqrt{\eta} ,
	\end{equation}
	so $\mathcal{F}^{\mathcal{K}}_{\mu\nu}=0$ if and only if $\mathcal{F}^{\mathcal{G}}_{\mu\nu}=0$. Hence the two integrability conditions characterize the same geometric obstruction. In summary, the directional quantity $\Delta_{n,\mu\nu}$ and the operator curvature $\mathcal{F}^{\mathcal{K}}_{\mu\nu}$ probe the same extension problem at different levels.

	\subsection{Cyclic protocols and endpoint holonomy}

	For a cyclic parameter protocol, the parameters return to their initial
	values, but the pathwise Hermitian representation need not return to its
	initial form.
	A proper transformation exists along the loop, but its
	endpoint may differ from its initial value by a unitary sewing
	transformation determined by the holonomy. The Hermitian Hamiltonian
	$\widetilde H=S_pHS_p^{-1}$ can therefore fail to close even when the
	quasi-Hermitian family is periodic. Such a nonperiodicity was observed in Ref.~\cite{JPAGW13} for a $\mathcal{PT}$-symmetric two-level system, where a closed parameter path in the original system was found to correspond to an open path in the mapped Hermitian system, but the underlying geometric mechanism was not identified. To see this, consider a closed loop $\mathcal{C}$ in parameter space, parametrized by, say, an angle $\theta$ with period $2\pi$. The evolution of the unitary factor $U$ in $S_{p} = U\sqrt{\eta}$ along $\mathcal{C}$ is governed by $\partial_\mu U = -U \mathcal{G}_\mu$, whose solution is given by the path-ordered exponential~\cite{Fishbane1981,Chepelev2002}
	\begin{equation}\label{eq:formula_013}
		U(\theta_f) = U(\theta_i) \, \mathcal{P} \exp\!\left(-\int_{\mathcal{C}} \mathcal{G}_\mu \,\mathrm{d}R^\mu\right) \equiv U(\theta_i) \, W(\mathcal{C}) ,
	\end{equation}
	where $\mathcal{P}$ denotes path ordering and $W(\mathcal{C})$ is the Wilson loop of the connection $\mathcal{G}_\mu$~\cite{Nakahara2003}. For a contractible loop $\mathcal C$ bounding a surface $\Sigma$, the non-Abelian Stokes theorem~\cite{Fishbane1981,Chepelev2002} relates $W(\mathcal C)$ to the curvature $\mathcal F^{\mathcal G}_{\mu\nu}$:
	\begin{equation}\label{eq:formula_014}
		W(\mathcal{C}) = \mathcal{P}_S \exp \left( -\frac{1}{2} \iint_{\Sigma} Q \mathcal{F}^{\mathcal{G}}_{\mu\nu} Q^{-1} d\Sigma^{\mu\nu} \right)
	\end{equation}
	where $Q$ is a unitary gauge bridge (Wilson line) performing parallel transport, $\mathcal{P}_S$ denotes surface ordering, and $\Sigma$ is a surface bounded by $\mathcal{C}$.
	
	If the curvature $\mathcal{F}^{\mathcal{G}}_{\mu\nu}$ is nonzero on $\Sigma$, the Wilson loop generically differs from the identity, $W(\mathcal{C}) \neq I$, unless the integral happens to vanish because of special quantization conditions. Now, the original quasi-Hermitian system is physically single-valued: the Hamiltonian $H$ and the metric $\eta$ (and hence $\sqrt{\eta}$) are periodic functions of the parameters. For a closed loop that returns to the same parameter point, we have $\sqrt{\eta}(\theta_f) = \sqrt{\eta}(\theta_i)$. The decomposition $S_{p} = U\sqrt{\eta}$ then gives
	\begin{equation}\label{eq:formula_015}
		S_{p}(\theta_f) = U(\theta_f) \sqrt{\eta}(\theta_f) = U(\theta_i) W(\mathcal{C}) \sqrt{\eta}(\theta_i) .
	\end{equation}
	The transformed Hermitian Hamiltonian at the endpoint is
	\begin{align}\label{eq:formula_016}
		&\tilde{H}(\theta_f) = S_{p}(\theta_f) H(\theta_f) S_{p}^{-1}(\theta_f)\notag \\
		&= U(\theta_i) W(\mathcal{C}) \sqrt{\eta}(\theta_i) H(\theta_i) \sqrt{\eta}(\theta_i)^{-1} W(\mathcal{C})^{-1} U(\theta_i)^{-1} \notag\\
		&= \bigl( U(\theta_i) W(\mathcal{C}) U(\theta_i)^{-1} \bigr) \, \tilde{H}(\theta_i) \, \bigl( U(\theta_i) W(\mathcal{C}) U(\theta_i)^{-1} \bigr)^{-1}.
	\end{align}
	After one cycle, the endpoint Hamiltonian is related to the initial one by the adjoint action of the holonomy:
	
	\begin{equation}\label{eq:formula_017}
		\begin{aligned}
			\widetilde H(\theta_f) =
			V(\mathcal C)
			\widetilde H(\theta_i)
			V(\mathcal C)^{-1},\\
			\qquad
			V(\mathcal C)
			=
			U(\theta_i)W(\mathcal C)U(\theta_i)^{-1}.
		\end{aligned}
	\end{equation}
	
	A nontrivial Wilson loop does not by itself imply a nonperiodic mapped Hamiltonian. Closure fails only when the adjoint action of $V(\mathcal C)$ on $\widetilde H(\theta_i)$ is nontrivial. If the holonomy is central, or more generally if $V(\mathcal C)$ commutes with $\widetilde H(\theta_i)$, the Hamiltonian remains periodic even though the proper frame has a monodromy. Nonzero curvature generically produces nontrivial holonomies and open Hermitian images, while special loops with trivial, central, commuting, or quantized holonomy may still close. This is the origin of the twisted boundary conditions in the Hermitian description.
	
	To connect the endpoint relation to a readout, consider a normalized, nondegenerate eigenstate $|u_n(0)\rangle$ of $\widetilde H(0)$ and a closed path $\mathcal C$ along which the level remains isolated. Adiabatic evolution then gives
	\begin{equation}
		\left|\psi_{\mathcal C}(T)\right\rangle
		\simeq
		\mathrm e^{\mathrm{i}\alpha_{n,\mathcal C}}
		V(\mathcal C)|u_n(0)\rangle,
		\label{eq:general-adiabatic-endpoint}
	\end{equation}
	where $\alpha_{n,\mathcal C}$ is the accumulated dynamical and geometric phase. The phase accounts for the evolution along the path, while the holonomy fixes the endpoint eigenstate orientation. A fixed self-adjoint observable $A_{\mathrm{ref}}$ in the mapped Hermitian space probes this orientation.

	A closely related geometric mismatch has recently been identified in the mixed-state Uhlmann phase of quasi-Hermitian systems, where the parameter-dependent metric induces an extra connection term modifying parallel transport and causing deviations from its Hermitian counterpart~\cite{hou2026theoryuhlmannphasequasihermitian}.
	
	%The remaining global issue on non-simply-connected parameter domains is discussed in Sec.~\ref{Sec.4}.
	
	\section{Topological obstruction}\label{Sec.4}
	
	When $\mathcal F^{\mathcal G}_{\mu\nu}=0$, proper frames exist locally. Their continuation around noncontractible loops determines whether these local frames extend to a globally single-valued one.
	
	For a closed path $\mathcal C$ based at $\boldsymbol R_0$,
	the Wilson loop is\cite{Simon1983,WilczekZee1984,Nakahara2003}
	\begin{equation}\label{eq:formula_018}
		W(\mathcal C)
		=
		\mathcal P\exp\left(
		-\oint_{\mathcal C}
		\mathcal G_\mu\,\mathrm dR^\mu
		\right).
	\end{equation}
	Writing the initial proper frame as
	$S_p(0)=U_0\sqrt{\eta(\boldsymbol R_0)}$,
	with $U_0$ unitary, its endpoint is
	$S_{p,\mathcal C}(T)
	=U_0W(\mathcal C)\sqrt{\eta(\boldsymbol R_0)}$.
	Thus, $W(\mathcal C)\neq I$ prevents the transported frame
	from returning to its initial value and rules out a globally
	single-valued proper frame.
	
	For a flat connection, the holonomy depends only on the
	based homotopy class of the loop. Contractible loops have
	trivial holonomy, whereas noncontractible loops may have
	nontrivial holonomy. We refer to the latter as a topological
	obstruction for the fixed structure $(H,\eta)$:
	it depends on both the metric-induced connection and the
	topology of the parameter domain.
	
	The endpoint action on operators depends on the sewing unitary $V(\mathcal C)=U_0W(\mathcal C)U_0^\dagger$. The mapped Hamiltonian closes precisely when
	\begin{equation}
		V(\mathcal C)\widetilde H(0)V^\dagger(\mathcal C)=\widetilde H(0),
		\qquad [V(\mathcal C),\widetilde H(0)]=0 .
		\label{eq:formula_019}
	\end{equation}
	A central holonomy leaves all operator images unchanged under conjugation, whereas a noncentral holonomy can open the Hermitian image when its adjoint action on the relevant operator is nontrivial.
	
	For a flat connection, a contractible loop returns the initial frame, whereas a noncontractible loop may rotate it. Starting from the same initial proper frame and isolated eigenstate, this rotation can be read out by measuring a fixed self-adjoint observable in the mapped Hermitian space:
	\begin{equation}
		\begin{aligned}
			\langle A_{\mathrm{ref}}\rangle_{T,\mathcal C}
			&\longrightarrow
			\langle u_n(0)|V^\dagger(\mathcal C)A_{\mathrm{ref}}V(\mathcal C)|u_n(0)\rangle,\\
			\Delta A_{\mathcal C}
			&=\langle u_n(0)|[V^\dagger(\mathcal C)A_{\mathrm{ref}}V(\mathcal C)-A_{\mathrm{ref}}]|u_n(0)\rangle .
		\end{aligned}
		\label{eq:general-reference-readout}
	\end{equation}
	The contrast depends on the chosen state and measurement axis. It is invariant under a common constant unitary change of basis applied to the state, holonomy, and observable.
	
	On a connected domain, flatness and trivial holonomy around all closed loops together characterize the existence of a global proper frame. Pathwise equivalence remains available throughout, with the endpoint action recording how the pathwise frames fit together.
	
	% ===== new =====
	Three cases can be distinguished. (i)~If $W(\mathcal C)=I$ for every closed path, a globally single-valued proper frame exists: the Hermitian image closes, the readout~\eqref{eq:general-reference-readout} returns the initial value, and the quasi-Hermitian and Hermitian geometric phases coincide for all loops. (ii)~If $W(\mathcal C)=\me^{\mi\theta}I$ is nontrivial but central, conjugation acts trivially on every operator, so the mapped Hamiltonian and observables remain periodic and the readout exhibits no contrast. The obstruction is confined to the proper frame and to the endpoint identification of cyclic representatives: the transported representative acquires an endpoint overlap $\me^{\mi\theta}$, and the gauge-invariant geometric phase of Sec.~II\,A acquires the endpoint contribution
	\begin{equation}\label{eq:central-endpoint-phase}
		\gamma_{n}=\theta+\oint\mathcal{A}_{n,\mu}\,\mathrm{d}R^{\mu}\pmod{2\pi},
	\end{equation}
	where $\theta$ is the phase of the central holonomy $W(\mathcal C)=\me^{\mi\theta}I$ (for a $\mathbb{Z}_2$ holonomy, $\theta=\pi$ for every loop in that class). This phase signature is physical only once an interferometric reference is specified; without one, all operator-level predictions are unchanged and the obstruction is a frame-level geometric classification. (iii)~If $W(\mathcal C)$ is noncentral and $[V(\mathcal C),\widetilde H(\theta_i)]\neq0$, the Hermitian image remains open and the endpoint state is rotated away from the initial eigenstate, changing the expectation value of $A_{\mathrm{ref}}$ and producing the contrast~\eqref{eq:general-reference-readout} demonstrated in Example~3 and, for a curved connection, in Example~4. Only here does the obstruction yield an unconditional operator-level contrast.
	
	This contrast differs from ordinary Hermitian path dependence. In a Hermitian system, an adiabatic, nondegenerate cycle returning the Hamiltonian to its initial value returns the state to the same eigenstate up to a phase, so every fixed-observable expectation value is unchanged. Here the same cyclic protocol returns $(H,\eta)$, and hence the spectrum, to its initial value, yet the holonomy rotates the endpoint state away from the initial eigenstate, changing $\langle A_{\mathrm{ref}}\rangle$. The contrast is therefore fixed by the holonomy rather than by protocol details: for a flat connection it depends only on the based homotopy class of the loop. It cannot be attributed to a choice of representation, nor reproduced by any globally single-valued proper Hermitian description of the same dynamics.
	
	\section{Examples}\label{Sec.5}
	
	We now illustrate the two types of obstructions with four examples. The first example exhibits neither obstruction; the second shows a geometric obstruction; the third example demonstrates a topological obstruction. The fourth connects the same geometry to a two-sublattice chain. Example 3 exhibits the obstruction through endpoint readout differences, while Example 4 illustrates operational signatures through endpoint readout and relative-phase protocols. Examples ~3 and ~4 also suggest possible experimental protocols. In all four examples, the metric is treated as part of the specified quasi-Hermitian structure $(H,\eta)$ and held fixed.
	
	\subsection{Example 1: No obstruction}
	
	We first examine a system free from both obstructions. Consider the quasi-Hermitian Hamiltonian
	\begin{equation}\label{eq:formula_020}
		H_0(\boldsymbol{R}) = B(\boldsymbol{R})\,\sigma_x + \mathrm{i}\gamma(\boldsymbol{R})\,\sigma_z,
	\end{equation}
	where $B(\boldsymbol{R})$ and $\gamma(\boldsymbol{R})$ are real functions satisfying $B(\boldsymbol{R})>|\gamma(\boldsymbol{R})|$, so that the eigenvalues are real. We choose the metric
	\begin{equation}\label{eq:formula_021}
		\eta_0(\boldsymbol{R}) = I + \frac{\gamma(\boldsymbol{R})}{B(\boldsymbol{R})}\,\sigma_y .
	\end{equation}
	Define the ratio $v = \gamma/B$. Since $B^2-\gamma^2>0$, one has $|v|<1$. The square root of the metric is
	\begin{equation}\label{eq:formula_022}
		\sqrt{\eta_0} = (1-v^2)^{1/4}\, \exp\!\Bigl(\frac12 \operatorname{arctanh}v\; \sigma_y\Bigr).
	\end{equation}
	A direct calculation shows that $\partial_\mu\sqrt{\eta_0}$ commutes with $\sqrt{\eta_0}^{-1}$ for any parameter direction $\mu$, because both are functions of $v$ and $\sigma_y$. Consequently, the gauge connection $\mathcal{G}_\mu = \frac12[\partial_\mu\sqrt{\eta_0},\sqrt{\eta_0}^{-1}]$ vanishes identically. Hence the curvature $\mathcal{F}_{\mu\nu}^{\mathcal{G}}$ vanishes automatically, and the Wilson loop for any closed loop $\mathcal{C}$ is the identity. Therefore neither a geometric nor a topological obstruction is present.
	
	We may take $S_p=\sqrt{\eta_0}$ (i.e., $U=I$), which is self-adjoint. This $S_p$ is smooth and globally single-valued. The mapped Hermitian Hamiltonian is
	\begin{equation}\label{eq:formula_023}
		\tilde{H}_0 = S_p H_0 S_p^{-1} = \sqrt{B^2 - \gamma^2}\;\sigma_x,
	\end{equation}
	which is indeed Hermitian. Here $S_p=\sqrt{\eta_0}$ is proper in every parameter direction, so Eq.~\eqref{eq:vanishing_curvature_difference} gives $\Delta_{n,\mu\nu}=0$ and $\widetilde{\mathcal F}_{n,\mu\nu}=\mathcal F_{n,\mu\nu}$. The pathwise transformations therefore combine into a single globally defined proper frame.

	\subsection{Example 2: Geometric obstruction}
	
	\begin{figure*}
		\centering
		\includegraphics[width=0.85\textwidth]{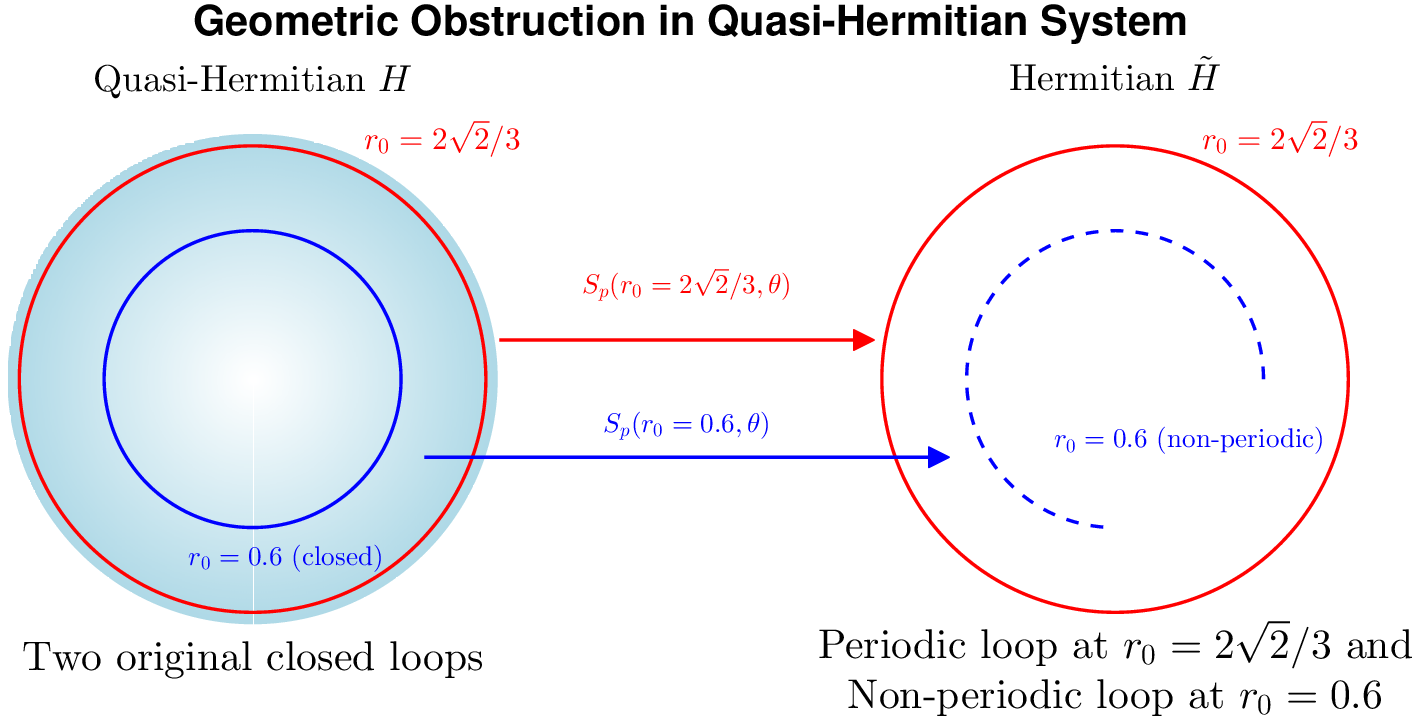}
		\caption{Geometric obstruction. (Left) Closed circular protocols at $r_0=2\sqrt{2}/3$ (red) and $r_0=0.6$ (blue). A pathwise proper transformation exists along either circular trajectory. (Right) The red Hermitian image closes, while the blue image remains open. The obstruction concerns the extension of these circle-by-circle transformations to a common proper frame over the disk.}
		\label{fig:geometric_obstruction}
	\end{figure*}
	
	Next, we consider a system that exhibits a geometric obstruction. The parameter space is a disk parametrized by polar coordinates $(r,\theta)$ with $0 \le r < 1$ and $0 \le \theta < 2\pi$.
	We choose a fixed parameter $\alpha$ with $\sin\alpha\neq0$ and introduce an $r$-dependent angle $\beta_r$ through $\sin\beta_r=r\sin\alpha$ and $\cos\beta_r=\sqrt{1-r^2\sin^2\alpha}$, which is strictly positive throughout the parameter domain.
	Define $\gamma_r = 1/\sqrt{1-r^2}$ and $g(r)=(\gamma_r-1)/2$. The Hamiltonian is given by
	\begin{align}\label{eq:formula_024}
		H(r,\theta) = H_x\sigma_x + H_y\sigma_y + H_z\sigma_z,
	\end{align}
	where
	
	\begin{align}\label{eq:formula_025}
		H_z &= \gamma_r\cos\beta_r
		+ \mathrm{i}\,\gamma_r r\sin\beta_r\sin(2\theta),\notag\\
		H_x &= \sin\beta_r\left(\frac{\gamma_r+1}{2}\cos\theta
		- \frac{\gamma_r-1}{2}\cos(3\theta)\right)
		\notag\\
		&\quad - \mathrm{i}\gamma_r r\cos\beta_r\sin\theta,\notag\\
		H_y &= \sin\beta_r\left(-\frac{\gamma_r+1}{2}\sin\theta
		- \frac{\gamma_r-1}{2}\sin(3\theta)\right)
		\notag\\
		&\quad + \mathrm{i} \gamma_r r\cos\beta_r\cos\theta.
	\end{align}
	
	We take the metric
	\begin{equation}\label{eq:formula_026}
		\eta(r,\theta) = I + r\cos\theta\,\sigma_x + r\sin\theta\,\sigma_y .
	\end{equation}
	One can verify directly that $\eta H=H^\dagger\eta$
	and $H^2=I$, giving the spectrum $E_\pm=\pm1$. At $r=0$,
	one has $\sin\beta_r=0$, $\cos\beta_r=1$, and $\gamma_r=1$, so
	$H(0,\theta)=\sigma_z$ is independent of $\theta$. Furthermore, writing $x=r\cos\theta$ and $y=r\sin\theta$, one finds that both the ratio $(\gamma_r-1)/r^2=\gamma_r^2/(\gamma_r+1)$ and $\cos\beta_r$ are smooth functions of $r^2=x^2+y^2$. Together with $\sin\beta_r=r\sin\alpha$, this shows that all components of $H$ are smooth functions of $x$ and $y$ at the origin. Thus the quasi-Hermitian structure $(H,\eta)$ is smooth through the disk.
	
	To analyze the connection $\mathcal{G}_\mu$, we write $\eta = I + r V$ with $V = \cos\theta\sigma_x + \sin\theta\sigma_y$, and express its square root and inverse in the form
	\begin{equation}\label{eq:formula_027}
		\sqrt{\eta} = a I + b V,\quad \sqrt{\eta}^{-1} = c I + d V,
	\end{equation}
	where $a,b,c,d$ are functions of $r$ satisfying $a^2+b^2=1$, $2ab=r$, $ac+bd=1$, and $bc+ad=0$. Solving these equations gives
	$a = \frac{\sqrt{1+r}+\sqrt{1-r}}{2}$, $b= \frac{\sqrt{1+r}-\sqrt{1-r}}{2}$, $c= \frac{1/\sqrt{1+r}+1/\sqrt{1-r}}{2}$, $d = \frac{1/\sqrt{1+r}-1/\sqrt{1-r}}{2}$.
	The components of $\mathcal{G}_\mu$ are
	
	\begin{align}\label{eq:formula_028}
		\mathcal{G}_\theta
		&=
		\frac12
		\bigl[
		\partial_\theta\sqrt{\eta},
		\sqrt{\eta}^{-1}
		\bigr]
		=
		-\mathrm{i}\,bd\,\sigma_z
		\notag\\
		&=
		\mathrm{i}
		\frac{r^2}
		{2\left(1-r^2+\sqrt{1-r^2}\right)}
		\sigma_z
		\neq0
		\qquad (r\neq0),
		\notag\\
		\mathcal{G}_r
		&=
		\frac12
		\bigl[
		\partial_r\sqrt{\eta},
		\sqrt{\eta}^{-1}
		\bigr]
		=
		0.
	\end{align}
	
	The curvature is therefore
	
	\begin{align}\label{eq:formula_029}
		\mathcal{F}^{\mathcal G}_{\theta r}
		&=
		\partial_\theta\mathcal G_r
		-
		\partial_r\mathcal G_\theta
		-
		[\mathcal G_\theta,\mathcal G_r]
		\notag\\
		&=
		\mathrm{i}\,\partial_r(bd)\,\sigma_z
		\notag\\
		&=
		-\mathrm{i}
		\frac{r}
		{2(1-r^2)^{3/2}}
		\sigma_z
		\neq0
		\qquad (r\neq0).
	\end{align}
	
	Thus a geometric obstruction exists: the connection is not flat, and the pathwise proper transformations cannot be extended to a single transformation that is proper throughout the disk.
	
	The nonzero curvature does not obstruct proper transport along a fixed circle. We first consider the coefficient-one ansatz below, whose properness selects a particular radius. For a fixed radius $r_0$, consider the ansatz $S_{p}(r_0,\theta) = \me^{-\mi\theta\sigma_z}\sqrt{\eta(r_0,\theta)}$.
	For notational convenience, let $\beta_0\equiv\beta_{r_0}$.
	To determine the required $r_0$, we first compute the Berry connections. For the mapped Hermitian Hamiltonian
	
	\begin{equation}\label{eq:formula_030}
		\tilde{H}(\theta) = S_{p} H S_{p}^{-1}
		= \sin\beta_0(\cos\theta\,\sigma_x + \sin\theta\,\sigma_y)
		+ \cos\beta_0\,\sigma_z,
	\end{equation}
	
	the low-lying ($E=-1$) eigenstate is
	
	\begin{equation}\label{eq:formula_031}
		|\psi_-\rangle =
		\begin{pmatrix}
			\sin(\beta_0/2)\,\me^{-\mi\theta} \\
			-\cos(\beta_0/2)
		\end{pmatrix},
	\end{equation}
	
	and its Berry connection is
	
	\begin{equation}\label{eq:formula_032}
		\mathcal{A}^\text{H}_\theta
		= \mi\langle\psi_-|\partial_\theta\psi_-\rangle
		= \frac{1 - \cos\beta_0}{2}.
	\end{equation}
	
	For the quasi-Hermitian system, the corresponding eigenstate is $|\Psi_-\rangle = \sqrt{\eta}^{-1}\me^{\mi\theta\sigma_z}|\psi_-\rangle$, and the quasi-Hermitian Berry connection (for the same level) is
	
	\begin{equation}\label{eq:formula_033}
		\mathcal{A}^\text{NH}_\theta
		= \mathcal{A}^\text{H}_\theta
		+ \cos\beta_0
		- g(r_0)\cos\beta_0 .
	\end{equation}
	
	Combining this expression with the directional equality
	$\mathcal A^{\mathrm{NH}}_\theta=\mathcal A^{\mathrm H}_\theta$
	and using $\cos\beta_0>0$, we obtain $g(r_0)=1$. The same condition follows directly by inserting
	$S_p(r_0,\theta)=\me^{-\mi\theta\sigma_z}\sqrt{\eta(r_0,\theta)}$
	into
	$S_p^\dagger\partial_\theta S_p=\frac12\partial_\theta\eta$.
	The coefficient-one ansatz is therefore proper only at $r_0=2\sqrt{2}/3$. This restriction is specific to that ansatz; pathwise proper solutions still exist along circles of other radii.
	
	For this special radius, the mapped Hamiltonian $\tilde{H}(\theta)$ is periodic with period $2\pi$. At a general fixed radius, the pathwise proper solution takes the form $S_p^{\mathrm{new}}(r_0,\theta) = \me^{-\mi g(r_0)\theta\sigma_z}\sqrt{\eta(r_0,\theta)}$, where $g(r_0)$ is determined by the connection. The resulting Hermitian Hamiltonian becomes
	
	\begin{align}\label{eq:formula_034}
		&\tilde{H}_{\text{new}}(\theta)
		= \me^{\mi(1-g(r_0))\theta\sigma_z}
		\tilde{H}(\theta)
		\me^{-\mi(1-g(r_0))\theta\sigma_z}
		\notag \\
		&= \sin\beta_0
		\Bigl[
		\cos\bigl((2g(r_0)-1)\theta\bigr)\,\sigma_x
		+
		\sin\bigl((2g(r_0)-1)\theta\bigr)\,\sigma_y
		\Bigr]
		\notag\\
		&\quad + \cos\beta_0\,\sigma_z.
	\end{align}
	
	The endpoint rotation is directly visible in the mapped Hermitian Hamiltonian. Define $k_0\equiv2g(r_0)-1$. Then Eq.~\eqref{eq:formula_034} gives
	\begin{equation}
		\begin{aligned}
			&\widetilde H_{\mathrm{new}}(0)=\sin\beta_0\,\sigma_x+\cos\beta_0\,\sigma_z,
			\\
			\widetilde H_{\mathrm{new}}(2\pi)&=\sin\beta_0[\cos(2\pi k_0)\sigma_x+\sin(2\pi k_0)\sigma_y]\\
			&+\cos\beta_0\,\sigma_z .
			\label{eq:ex2-final-H}
		\end{aligned}
	\end{equation}
	In the standard parameter space, $\theta$ runs from $0$ to $2\pi$.
	For the nonzero circular trajectories considered here, $\sin\beta_0=r_0\sin\alpha\neq0$. Therefore, for the mapped Hamiltonian to be single-valued, $(2g(r_0)-1)$ must be an integer.
	However, for the transformation $S_p^{\mathrm{new}}(r_0,\theta)$ itself to be single-valued (a necessary condition for a global proper transformation $S_p$), its phase factor $\me^{-\mi g(r_0)\theta\sigma_z}$ must return to identity after a full cycle, which requires $g(r_0)$ to be an integer.
	Requiring both the proper transformation and its Hermitian image to be single-valued gives, using $2g(r_0)-1=\gamma_{r_0}-2$ and $\gamma_{r_0}>1$, the condition $\gamma_{r_0}=3,5,7,\ldots$.
	The smallest nontrivial solution is $\gamma_{r_0}=3$, which gives $r_0=2\sqrt{2}/3$. Figure~\ref{fig:geometric_obstruction} illustrates this: the red loop with $r_0=2\sqrt{2}/3$ ($\gamma_{r_0}=3$) maps to a closed Hermitian loop, while the blue loop with non-integer $g(r_0)$ (e.g., $r_0=0.6$) maps to an open path.
	
	For $k_0\notin\mathbb Z$, the endpoint Hamiltonian generally differs from its initial value. In the mapped Hermitian space, choose the fixed observable $A_{\mathrm{ref}}^{(r_0)}\equiv\widetilde H_{\mathrm{new}}(0)$ and prepare its positive-energy eigenstate. Adiabatic evolution gives
	\begin{equation}
		\left\langle A_{\mathrm{ref}}^{(r_0)}\right\rangle_{2\pi}
		\longrightarrow
		\cos^2\beta_0+\sin^2\beta_0\cos(2\pi k_0),
		\label{eq:ex2-final-expectation}
	\end{equation}
	whereas $\left\langle A_{\mathrm{ref}}^{(r_0)}\right\rangle_0=1$. The endpoint contrast therefore measures the rotation relative to the initial Hermitian axis. At the special radius $r_0=2\sqrt2/3$, one has $k_0=1$, so both the Hermitian image and the reference expectation value close. This phenomenon was observed in Ref.~\cite{JPAGW13}, where a closed parameter loop in the $\mathcal{PT}$-symmetric system was found to map to an open path in the equivalent Hermitian system. Our analysis identifies the curvature of $\mathcal{G}_\mu$ as the origin of this effect. When the curvature vanishes and the parameter space is simply connected, the connection $\mathcal{G}_\mu$ becomes pure gauge, and a global $S_p$ can be chosen such that the mapped Hamiltonian inherits the periodicity of the original system. In the presence of curvature, however, such a global choice is impossible, and the geometric obstruction manifests as a twisted boundary condition for the Hermitian image.

	\subsection{Example 3: Topological obstruction}
	\begin{figure*}
		\centering
		\includegraphics[width=0.85\textwidth]{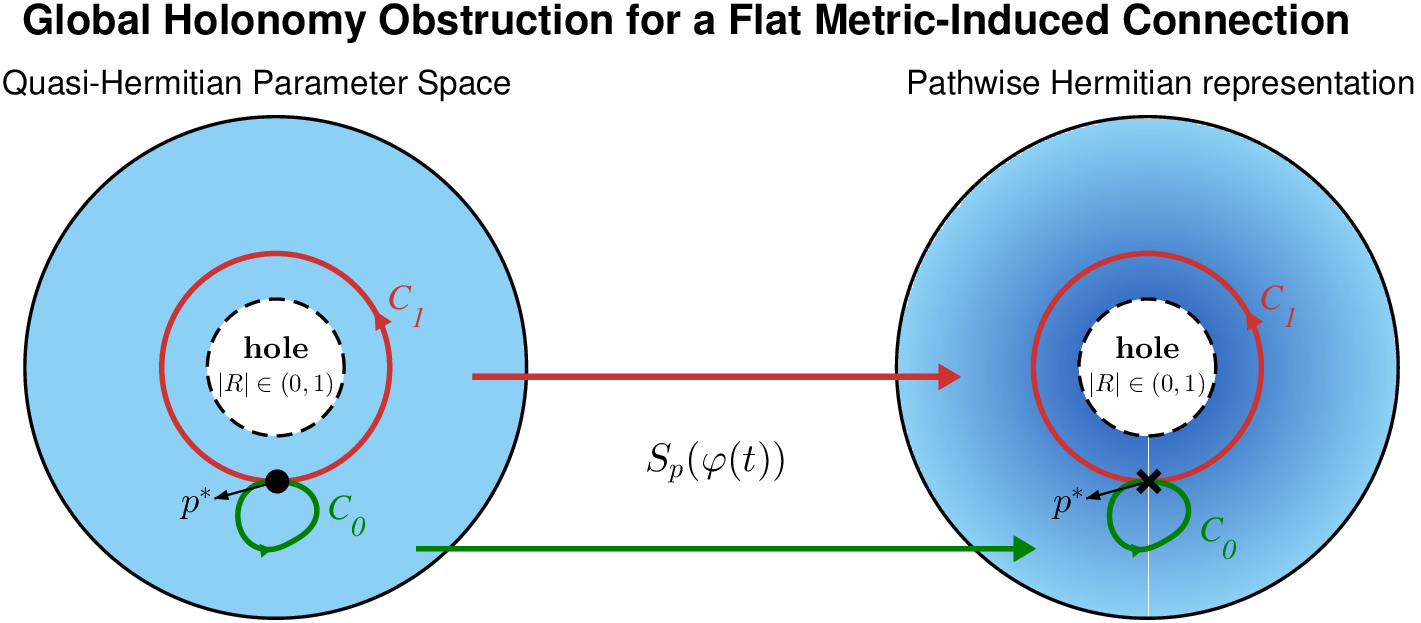}
		\caption{
			Topological obstruction for a flat metric-induced connection.
			(Left) Both quasi-Hermitian parameter paths are closed at the
			common base point $P^*$ (black dot). The green loop $\mathcal C_0$
			is contractible, whereas the red loop $\mathcal C_1$ winds once
			around the hole.
			(Right) Under proper transport from the same initial frame,
			the Hermitian image of $\mathcal C_0$ closes, while that of
			$\mathcal C_1$ does not, owing to the noncentral holonomy
			$W_1=\exp(-\mathrm{i}\pi\sigma_z/4)$.
			The cross marks the endpoint mismatch of the latter.
			In the mapped Hermitian space, for the common initial state $|+x\rangle$ and fixed observable
			$A_{\mathrm{ref}}=\widetilde H(0)=\sigma_x$, the initial
			expectation value is $1$, while the adiabatic endpoint values
			approach $1$ along $\mathcal C_0$ and $0$ along $\mathcal C_1$.
		}
		\label{fig:topological_obstruction}
	\end{figure*}
	A flat connection with nontrivial holonomy is realized by the following two-level model. The parameter space is an annulus,\[
	\mathcal{M}
	=
	\{(R,\phi)\,|\,1<R<2,\ 0\leq\phi<2\pi\}.\]
	We introduce the Hermitian matrix 
	\begin{equation} \label{Q}
		Q(\phi)=\cos\phi\,\sigma_x+\sin\phi\,\sigma_y
	\end{equation}
	which satisfies $Q^\dagger(\phi)=Q(\phi),  Q^2(\phi)=I$.
	The metric is chosen as
	\begin{equation} \label{eta}
		\eta(\phi)=I+\frac35 Q(\phi).
	\end{equation}
	It is positive definite throughout the parameter domain.	
	
	The Hermitian matrix $h_0=\sigma_x$ sets the energy scale in the algebraic construction. The positive square root of the metric and its inverse are
	\begin{equation}
		\begin{aligned}
			&N(\phi)=\sqrt{\eta(\phi)}
			=\frac{1}{\sqrt{10}}\left(3I+Q(\phi)\right),\\
			&N^{-1}(\phi)=\frac{\sqrt{10}}{8}\left(3I-Q(\phi)\right).
		\end{aligned}
		\label{eq:ex3-metric-root}
	\end{equation}
	We then construct the quasi-Hermitian Hamiltonian
	\begin{equation}
		\begin{aligned}
			H(\phi)
			&=N^{-1}(\phi)h_0N(\phi)\\
			&=\frac18
			\left[
			(9-\cos2\phi)\sigma_x
			-\sin2\phi\,\sigma_y
			+6\mathrm{i}\sin\phi\,\sigma_z
			\right].
		\end{aligned}
		\label{eq:ex3-quasi-Hermitian-H}
	\end{equation}
	Because $N^\dagger=N$ and $\eta=N^2$, one has
	\begin{equation}
		H^\dagger\eta
		=
		N\sigma_xN
		=
		\eta H.
		\label{eq:ex3-quasi-Hermiticity}
	\end{equation}
	Thus, $H$ is quasi-Hermitian with respect to the specified metric $\eta$,
	and its spectrum is $E_\pm=\pm1$.
	Both $H$ and $\eta$ are $2\pi$-periodic, and in particular
	$H(2\pi)=H(0)=\sigma_x$.
	
	For a parameter path $(R(t),\phi(t))$, the quasi-Hermitian system is assumed
	to obey the metric-compatible dynamics
	\begin{equation}
		\mathrm{i}\partial_t|\Psi(t)\rangle
		=
		\left[
		H(\phi(t))
		-\frac{\mathrm{i}}{2}
		\eta^{-1}(\phi(t))
		\frac{\mathrm{d}\eta(\phi(t))}{\mathrm{d}t}
		\right]
		|\Psi(t)\rangle .
		\label{eq:ex3-quasi-Hermitian-dynamics}
	\end{equation}
	The specified structure $(H,\eta)$ determines proper transport along each parameter path.
	
	\paragraph{Proper transport and the flat metric-induced connection}
	
	We write the proper Dyson map as
	\begin{equation}
		S_p=UN,
		\qquad
		U^\dagger U=I,
		\label{eq:ex3-Sp-decomposition}
	\end{equation}
	so that $S_p^\dagger S_p=\eta$ is automatically satisfied.
	The proper condition along the angular direction is
	\begin{equation}
		(\partial_\phi S_p)^\dagger S_p
		=
		S_p^\dagger(\partial_\phi S_p)
		=
		\frac12\partial_\phi\eta .
		\label{eq:ex3-proper-condition}
	\end{equation}
	The corresponding equation for the unitary factor is
	\begin{equation}
		\partial_\phi U=-UG_\phi,
		\qquad
		G_\phi
		=
		\frac12
		[\partial_\phi N,N^{-1}] .
		\label{eq:ex3-unitary-transport}
	\end{equation}
	For later use, define
	\begin{equation}
		Q'(\phi)\equiv\partial_\phi Q(\phi)
		=
		-\sin\phi\,\sigma_x+\cos\phi\,\sigma_y .
		\label{eq:ex3-Q-prime}
	\end{equation}
	Using $[Q',Q]=-2\mathrm{i}\sigma_z$, one obtains
	\begin{equation}
		G_\phi
		=
		\frac1{16}[Q',3I-Q]
		=
		\frac{\mathrm{i}}8\sigma_z,
		\qquad
		G_R=0 .
		\label{eq:ex3-connection}
	\end{equation}
	The radial component vanishes because the metric is independent of $R$.
	Consequently, the curvature of the metric-induced connection is
	\begin{equation}
		F_{\phi R}^{G}
		\equiv
		\partial_\phi G_R
		-\partial_R G_\phi
		-[G_\phi,G_R]
		=0 .
		\label{eq:ex3-curvature}
	\end{equation}
	The connection is flat, so global compatibility is determined by its holonomy around the hole.
	
	Choose a common base point
	\begin{equation}
		P^*=(R_0,0),
		\qquad
		1<R_0<2 .
		\label{eq:ex3-base-point}
	\end{equation}
	For a path starting from $P^*$, let $\varphi(t)\in\mathbb{R}$ denote the
	continuous lift of the angular coordinate. Thus,
	$\phi(t)=\varphi(t)\ {\rm mod}\ 2\pi$ and $\varphi(0)=0$.
	We impose the same initial condition for every path,
	\begin{equation}
		U(0)=I,
		\qquad
		S_p(0)=N(0).
		\label{eq:ex3-initial-frame}
	\end{equation}
	The proper transport equation then gives
	\begin{equation}
		\begin{aligned}
			&U(t)
			=
			\exp\left[
			-\frac{\mathrm{i}\varphi(t)}8\sigma_z
			\right],
			\\
			&S_p(t)
			=
			\exp\left[
			-\frac{\mathrm{i}\varphi(t)}8\sigma_z
			\right]
			N(\phi(t)).
			\label{eq:ex3-pathwise-Sp}
		\end{aligned}
	\end{equation}
	Because $G_R=0$, this expression applies also to paths whose radial coordinate
	varies with time. These solutions share the same initial frame; their endpoints determine the global single-valuedness of $S_p$.
	
	\paragraph{Two closed paths with the same endpoints}
	
	Consider closed paths that start and end at the same point $P^*$.
	Let $w\in\mathbb{Z}$ denote the winding number of such a path around the
	hole. The continuous angular lift then satisfies
	$\varphi(T)=2\pi w$.
	Since $G_\phi$ is proportional to the fixed matrix $\sigma_z$, path ordering
	is trivial, and the Wilson loop is
	\begin{equation}
		\begin{aligned}
			W_w
			&=
			\exp\left[
			-\int_0^T G_\phi\dot{\varphi}(t)\,\mathrm{d}t
			\right]\\
			&=
			\exp\left[
			-\frac{\mathrm{i}\pi w}{4}\sigma_z
			\right],
			\qquad
			S_p(T)=W_wS_p(0).
		\end{aligned}
		\label{eq:ex3-Wilson-loop}
	\end{equation}
	
	As illustrated in Fig.~\ref{fig:topological_obstruction}, we now compare two closed paths with the same initial and final point.
	Path $\mathcal C_0$ does not wind around the hole, whereas path $\mathcal C_1$ winds once in the
	positive direction. Their winding numbers are therefore $w_0=0$ and
	$w_1=1$, respectively. The corresponding holonomies and endpoint frames are
	\begin{equation}
		\begin{aligned}
			W_0&=I,
			&S_{p,0}(T)&=S_p(0),\\
			W_1&=\exp\left(-\frac{\mathrm{i}\pi}{4}\sigma_z\right),
			&S_{p,1}(T)&=W_1S_p(0).
		\end{aligned}
		\label{eq:ex3-two-holonomies}
	\end{equation}
	The frame closes along $\mathcal C_0$ but not along $\mathcal C_1$ (Fig.~\ref{fig:topological_obstruction}), establishing the global obstruction on the annulus.
	Here $W_1$ is noncentral and does not commute with $\sigma_x$.
	Its adjoint action therefore changes the mapped Hermitian
	Hamiltonian after one winding.
	
	\paragraph{The proper Hermitian images of the two paths}
	
	Define the mapped state by
	\begin{equation}
		|\psi(t)\rangle=S_p(t)|\Psi(t)\rangle .
		\label{eq:ex3-mapped-state}
	\end{equation}
	The proper condition implies
	$S_p^{-1}\dot S_p=\frac12\eta^{-1}\dot\eta$.
	Therefore, the derivative term generated by the time-dependent similarity
	transformation cancels the metric-dependent term in
	Eq.~\eqref{eq:ex3-quasi-Hermitian-dynamics}, and the mapped state obeys
	\begin{equation}
		\begin{aligned}
			&\mathrm{i}\partial_t|\psi(t)\rangle
			=
			\widetilde H(t)|\psi(t)\rangle,
			\\
			&\widetilde H(t)
			=
			S_p(t)H(\phi(t))S_p^{-1}(t).
			\label{eq:ex3-Hermitian-dynamics}
		\end{aligned}
	\end{equation}
	Using Eqs.~\eqref{eq:ex3-quasi-Hermitian-H} and
	\eqref{eq:ex3-pathwise-Sp}, we obtain
	\begin{equation}
		\begin{aligned}
			\widetilde H(t)
			&=U(t)\sigma_xU^\dagger(t)\\
			&=
			\cos\frac{\varphi(t)}4\,\sigma_x
			+
			\sin\frac{\varphi(t)}4\,\sigma_y .
		\end{aligned}
		\label{eq:ex3-mapped-H}
	\end{equation}
	Both paths start from the same Hermitian Hamiltonian,
	\begin{equation}
		\widetilde H(0)=\sigma_x .
		\label{eq:ex3-initial-H}
	\end{equation}
	For path $\mathcal C_0$, $\varphi_0(T)=0$, whereas for path $\mathcal C_1$,
	$\varphi_1(T)=2\pi$. Consequently,
	\begin{equation}
		\begin{aligned}
			&\widetilde H_0(T)
			=
			\widetilde H(0)
			=
			\sigma_x,
			\\
			&\widetilde H_1(T)
			=
			W_1\widetilde H(0)W_1^\dagger
			=
			\sigma_y .
			\label{eq:ex3-endpoint-H}
		\end{aligned}
	\end{equation}
	Both paths return to the same pair $(H,\eta)$, while their proper Hermitian images close along $\mathcal C_0$ and remain open along $\mathcal C_1$. This difference reflects the holonomy around the hole.
	
	\paragraph{Observable consequence in a fixed reference frame}
	
	For the endpoint measurement, prepare the positive-energy eigenstate \(|+x\rangle\) of \(\widetilde H(0)\) in the common Hermitian reference space and use the same fixed measurement axis for both protocols.
	
	Along either path, the positive-energy instantaneous eigenstate of
	$\widetilde H(t)$ can be chosen as
	\begin{equation}
		|u_+(t)\rangle
		=
		U(t)|+x\rangle,
		\qquad
		\widetilde H(t)|u_+(t)\rangle=|u_+(t)\rangle .
		\label{eq:ex3-instantaneous-state}
	\end{equation}
	The energy gap is equal to $2$ throughout the evolution.
	When both paths are traversed sufficiently slowly, the adiabatic limit gives
	\begin{equation}
		|\psi_w(T)\rangle
		\simeq
		\mathrm{e}^{\mathrm{i}\alpha_w}W_w|+x\rangle ,
		\label{eq:ex3-adiabatic-final-state}
	\end{equation}
	where $\alpha_w$ denotes the total dynamical and geometric phase accumulated
	along the path with winding number $w$. In Eq.~\eqref{eq:ex3-adiabatic-final-state}, $W_w$ determines the endpoint orientation of the adiabatically followed state.
	The fixed observable in this Hermitian space is
	\begin{equation}
		A_{\mathrm{ref}}
		=
		\sigma_x
		=
		\widetilde H(0).
		\label{eq:ex3-reference-observable}
	\end{equation}
	Keeping this measurement axis fixed, and using
	$\langle+x|\sigma_x|+x\rangle=1$, the initial reference expectation value is
	\begin{equation}
		\langle A_{\mathrm{ref}}\rangle_0=1 .
		\label{eq:ex3-initial-expectation}
	\end{equation}
	For a path with winding number $w$, Eq.~\eqref{eq:ex3-adiabatic-final-state}
	gives
	\begin{equation}
		\begin{aligned}
			\langle A_{\mathrm{ref}}\rangle_{T,w}
			&\longrightarrow
			\langle+x|W_w^\dagger\sigma_xW_w|+x\rangle\\
			&=
			\cos\frac{\pi w}{2}.
		\end{aligned}
		\label{eq:ex3-reference-expectation}
	\end{equation}
	The arrow indicates the adiabatic limit; at finite driving speed,
	nonadiabatic corrections may occur.
	
	In particular, path $\mathcal C_0$ gives
	\begin{equation}
		|\psi_0(T)\rangle
		\simeq
		\mathrm{e}^{\mathrm{i}\alpha_0}|+x\rangle ,
		\label{eq:ex3-final-state-b}
	\end{equation}
	whereas path $\mathcal C_1$ gives
	\begin{equation}
		\begin{aligned}
			&|\psi_1(T)\rangle
			\simeq
			\mathrm{e}^{\mathrm{i}\alpha_1}
			W_1|+x\rangle
			=
			\mathrm{e}^{\mathrm{i}(\alpha_1-\pi/4)}
			|+y\rangle ,
			\\
			&|+y\rangle
			=
			\frac1{\sqrt2}
			\begin{pmatrix}
				1\\
				\mathrm{i}
			\end{pmatrix}.
		\end{aligned}
		\label{eq:ex3-final-state-a}
	\end{equation}
	Therefore, the same fixed reference observable gives
	\begin{equation}
		\begin{aligned}
			\langle A_{\mathrm{ref}}\rangle_{T,0}
			\longrightarrow
			1
			=
			\langle A_{\mathrm{ref}}\rangle_0,
			\\
			\langle A_{\mathrm{ref}}\rangle_{T,1}
			\longrightarrow
			0
			\neq
			\langle A_{\mathrm{ref}}\rangle_0 .
		\end{aligned}
		\label{eq:ex3-observable-contrast}
	\end{equation}
	The endpoints $\phi=0$ and $\phi=2\pi$ coincide on the annulus; their continuous lifts $\varphi_0(T)=0$ and $\varphi_1(T)=2\pi$ retain the winding information.
	
	In the adiabatic limit, the fixed-axis contrast depends only on the winding number and is preserved under smooth deformations within the same based homotopy class.
	
	\paragraph{Possible experimental protocol.}
	
	The preceding construction suggests a possible experimental protocol based
	on the quasi-Hermitian system itself. The system is prepared at the common
	base point $P^*=(R_0,0)$ in the positive-energy state of $H(0)$, normalized
	with respect to $\eta(0)$. The protocol requires control of $H(\phi)$, the
	positive metric $\eta(\phi)$, and the metric-compatible generator specified
	in Eq.~\eqref{eq:ex3-quasi-Hermitian-dynamics} along the prescribed path.
	
	Two protocols are performed with the same initial state and the same initial
	proper frame $U(0)=I$, $S_p(0)=N(0)$. The first path $\mathcal C_0$ has
	winding number $w_0=0$, whereas $\mathcal C_1$ winds once around the hole
	and has $w_1=1$. Their proper transports and endpoint frames are given in
	Eq.~\eqref{eq:ex3-two-holonomies}; both paths nevertheless return to the
	same quasi-Hermitian pair $(H,\eta)$.
	
	At the endpoint, the states are mapped to the common Hermitian reference
	space by the corresponding pathwise proper maps, and the fixed self-adjoint
	observable $A_{\mathrm{ref}}=\sigma_x$ defined in
	Eq.~\eqref{eq:ex3-reference-observable} is measured. The adiabatic endpoint
	signals are those given in Eqs.~\eqref{eq:ex3-reference-expectation} and
	\eqref{eq:ex3-observable-contrast}, namely $1$ and $0$ for
	$\mathcal C_0$ and $\mathcal C_1$, respectively. This observable is defined
	in the mapped Hermitian reference space and need not be identified with a
	generic observable of the original quasi-Hermitian space.
	
	A direct quasi-Hermitian implementation requires simultaneous control of
	the non-Hermitian generator, the positive metric, and the metric-compatible
	evolution. The feasibility of such control depends on the specific experimental
	platform and may involve dilation-based, ancilla-assisted, or post-selected
	implementations. Alternatively, direct implementation of the proper Hermitian
	Hamiltonian in Eq.~\eqref{eq:ex3-mapped-H} provides a simpler route for
	simulating the corresponding metric-compatible quasi-Hermitian protocol in a
	conventional Hermitian setting.
	
	\subsection{Example 4: A two-sublattice chain}
	\label{sec:ex4_chain}
	
	A two-sublattice chain connects the curvature obstruction to the
	band structure of a quasi-Hermitian extension of the
	Su--Schrieffer--Heeger (SSH) and Rice--Mele (RM) models
	(hereafter SSH--RM model)
	\cite{Rice1982,Asboth2016,Kumar2025,Ruzicka2016},
	with readout protocols defined in its proper Hermitian images.
	With lattice spacing and $\hbar$ set to unity, let $\kappa$ denote
	the dimensionless Bloch momentum and $\xi$ a real, dimensionless
	control parameter. Define
	\begin{equation}
		\begin{aligned}
			&\mathsf X(\kappa)
			=\cos\kappa\,\sigma_x+\sin\kappa\,\sigma_y,
			\\
			&\mathsf Y(\kappa)
			=-\sin\kappa\,\sigma_x+\cos\kappa\,\sigma_y.
		\end{aligned}
		\label{ex4:axes}
	\end{equation}
	For a fixed hopping scale $J_{\mathrm{lat}}>0$ and a fixed real
	staggered-potential parameter $m_{\mathrm{lat}}\ne0$, specify
	\begin{equation}
		\begin{aligned}
			&\mathsf H_{\mathrm{lat}}(\kappa,\xi)
			\\
			&\quad=
			\frac{
				(J_{\mathrm{lat}}-\mathrm{i}m_{\mathrm{lat}}\xi)
				\mathsf X(\kappa)
				+(m_{\mathrm{lat}}+\mathrm{i}J_{\mathrm{lat}}\xi)\sigma_z
			}{\sqrt{1-\xi^2}},
			\\
			&\eta_{\mathrm{lat}}(\kappa,\xi)
			=I+\xi\mathsf Y(\kappa).
		\end{aligned}
		\label{ex4:model}
	\end{equation}
	At $\xi=0$, this reduces to the zero-intracell-hopping limit of a
	Rice--Mele chain with staggered onsite potential $m_{\mathrm{lat}}$
	\cite{Rice1982,Asboth2016}.
	The metric eigenvalues $1\pm\xi$ are positive on the cylindrical
	control space
	$\mathcal M_{\mathrm{lat}}=S^1_\kappa\times(-1,1)_\xi$,
	where $\kappa\sim\kappa+2\pi$.
	Under periodic boundary conditions this is a two-sublattice Bloch
	Hamiltonian with intercell hopping
	$(J_{\mathrm{lat}}-\mathrm{i}m_{\mathrm{lat}}\xi)/\sqrt{1-\xi^2}$
	in both directions and opposite onsite terms
	$\pm(m_{\mathrm{lat}}+\mathrm{i}J_{\mathrm{lat}}\xi)/\sqrt{1-\xi^2}$.
	It satisfies
	$\mathsf H_{\mathrm{lat}}^\dagger\eta_{\mathrm{lat}}
	=\eta_{\mathrm{lat}}\mathsf H_{\mathrm{lat}}$
	and has constant energies $E_\pm=\pm\varepsilon_{\mathrm{lat}}$,
	with
	$\varepsilon_{\mathrm{lat}}
	=\sqrt{J_{\mathrm{lat}}^2+m_{\mathrm{lat}}^2}$.
	The dynamics is the metric-compatible one specified above.
	
	\begin{figure}
		\centering
		\includegraphics[width=0.45\textwidth]{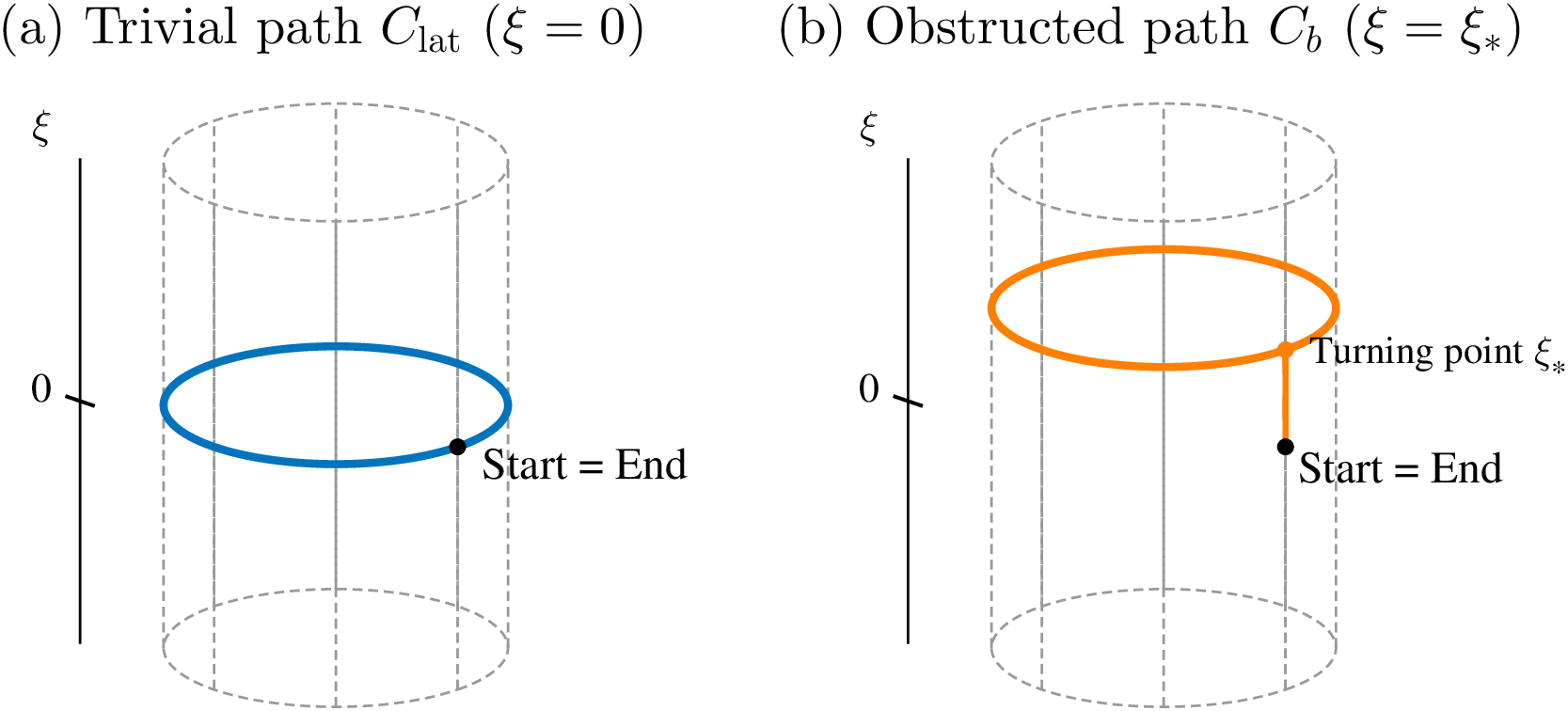}
		\caption{
			Geometric obstruction on the cylindrical control space
			$\mathcal M_{\mathrm{lat}}=S^1_\kappa\times(-1,1)_\xi$.
			(Left) The reference path $\mathcal C_{\mathrm{lat}}$
			at $\xi=0$ (blue closed loop).
			(Right) The path $\mathcal C_{\mathrm b}$ (orange)
			climbs to $\xi_*$, winds once around the cylinder,
			and returns, accumulating the holonomy angle
			$\zeta_*=2\pi f_{\mathrm{lat}}(\xi_*)$,
			with $f_{\mathrm{lat}}$ defined in
			Eq.~\eqref{ex4:connection}.
			For $m_{\mathrm{lat}}/J_{\mathrm{lat}}=1/2$
			and $\xi_*=3/5$, the fixed-reference endpoint
			expectations are $-1$ and $-1/5$.
		}
		\label{fig:cylindrical_obstruction}
	\end{figure}

	\paragraph{Proper transport and the geometric obstruction}
	Writing $\mathsf D=\sqrt{\eta_{\mathrm{lat}}}
	=\ell_+I+\ell_-\mathsf Y$, with
	$\ell_\pm=(\sqrt{1+\xi}\pm\sqrt{1-\xi})/2$,
	we denote the metric-induced connection by
	$G^{\mathrm{lat}}_\mu$, $\mu\in\{\kappa,\xi\}$,
	and its curvature by $F^{G,\mathrm{lat}}_{\kappa\xi}$.
	Using $\partial_\kappa\mathsf Y=-\mathsf X$ and
	$[\mathsf X,\mathsf Y]=2\mathrm{i}\sigma_z$ gives
	\begin{equation}
		\begin{aligned}
			G^{\mathrm{lat}}_\kappa
			&=\tfrac12[\partial_\kappa\mathsf D,\mathsf D^{-1}]
			=\mathrm{i}f_{\mathrm{lat}}(\xi)\sigma_z,
			\\
			G^{\mathrm{lat}}_\xi&=0,
			\\
			f_{\mathrm{lat}}(\xi)
			&=\tfrac12[(1-\xi^2)^{-1/2}-1],
			\\
			F^{G,\mathrm{lat}}_{\kappa\xi}
			&=-\frac{\mathrm{i}\xi\sigma_z}
			{2(1-\xi^2)^{3/2}}.
		\end{aligned}
		\label{ex4:connection}
	\end{equation}
	Thus a common proper frame is locally obstructed at $\xi\ne0$.
	Along a prescribed path $(\kappa(t),\xi(t))$, with
	$t\in[0,T]$ and $T$ the protocol duration, let
	$U_{\mathrm{lat}}$ be the unitary factor of the proper map
	$S_{p,\mathrm{lat}}$.
	With the common initial condition $U_{\mathrm{lat}}(0)=I$, set
	\begin{equation}
		\begin{aligned}
			&S_{p,\mathrm{lat}}=U_{\mathrm{lat}}\mathsf D,
			\qquad
			U_{\mathrm{lat}}
			=e^{-\mathrm{i}\zeta_{\mathrm{lat}}\sigma_z},
			\\
			&\zeta_{\mathrm{lat}}(t)
			=\int_0^t
			f_{\mathrm{lat}}(\xi(t'))\dot\kappa(t')\,dt'.
		\end{aligned}
		\label{ex4:proper}
	\end{equation}
	Here $\kappa(t)$ denotes the continuous angular lift,
	$\zeta_{\mathrm{lat}}(t)$ is the accumulated rotation parameter,
	and a dot denotes a time derivative.
	This map obeys
	$S_{p,\mathrm{lat}}^\dagger\dot S_{p,\mathrm{lat}}
	=\dot S_{p,\mathrm{lat}}^\dagger S_{p,\mathrm{lat}}
	=\dot\eta_{\mathrm{lat}}/2$.
	For a state $|\Psi\rangle$ in the original quasi-Hermitian space,
	the mapped state $|\psi\rangle=S_{p,\mathrm{lat}}|\Psi\rangle$
	therefore evolves with the Hermitian Hamiltonian
	\begin{equation}
		\begin{aligned}
			\widetilde{\mathsf H}_{\mathrm{lat}}
			&=S_{p,\mathrm{lat}}\mathsf H_{\mathrm{lat}}
			S_{p,\mathrm{lat}}^{-1}
			\\
			&=J_{\mathrm{lat}}
			[\cos(\kappa+2\zeta_{\mathrm{lat}})\sigma_x
			\\
			&\qquad
			+\sin(\kappa+2\zeta_{\mathrm{lat}})\sigma_y]
			+m_{\mathrm{lat}}\sigma_z.
		\end{aligned}
		\label{ex4:drive}
	\end{equation}
	This proper Hermitian Hamiltonian specifies the driving protocol
	for each path.
	
	\paragraph{Two paths and corresponding holonomy}
	As shown in Fig.~\ref{fig:cylindrical_obstruction}, consider two closed protocols based at $(0,0)$,
	\begin{equation}
		\begin{split}
			\mathcal C_{\mathrm{lat}}:\;&
			(0,0)\longrightarrow(2\pi,0),
			\\
			\mathcal C_{\mathrm b}:\;&
			(0,0)\longrightarrow(0,\xi_*)
			\\
			&\longrightarrow(2\pi,\xi_*)
			\longrightarrow(2\pi,0),
		\end{split}
		\label{ex4:paths}
	\end{equation}
	where $\xi_*$ is the fixed excursion height, with $0<\xi_*<1$.
	Both paths wind once around the cylinder.
	Their proper-transport holonomies are
	\begin{equation}
		W_{\mathrm{lat}}=I,
		\qquad
		W_{\mathrm b}=e^{-\mathrm{i}\zeta_*\sigma_z},
		\qquad
		\zeta_*=2\pi f_{\mathrm{lat}}(\xi_*).
		\label{ex4:holonomy}
	\end{equation}
	Their difference is caused by curvature between the paths:
	the composite loop $\mathcal C_{\mathrm b}$ followed by reversed
	$\mathcal C_{\mathrm{lat}}$ is contractible.
	This differs from the flat-connection obstruction in Example 3.
	
	\paragraph{Observable consequence in a fixed reference frame}
	Prepare the normalized lower eigenstate $|u_0\rangle$ of
	$\widetilde{\mathsf H}_{\mathrm{lat}}(0)$, with eigenvalue
	$-\varepsilon_{\mathrm{lat}}$, and traverse both paths adiabatically.
	Let $|\psi_\nu(t)\rangle$ denote the mapped state along
	$\mathcal C_\nu$, where $\nu\in\{\mathrm{lat},\mathrm b\}$.
	The endpoint states satisfy
	\begin{equation}
		\begin{aligned}
			|\psi_{\mathrm{lat}}(T)\rangle
			&\simeq e^{\mathrm{i}\alpha_{\mathrm{lat}}}|u_0\rangle,
			\\
			|\psi_{\mathrm b}(T)\rangle
			&\simeq e^{\mathrm{i}\alpha_{\mathrm b}}
			W_{\mathrm b}|u_0\rangle.
		\end{aligned}
		\label{ex4:endstates}
	\end{equation}
	Here $\alpha_\nu$ is the total accumulated dynamical and geometric
	phase along branch $\nu$.
	To read out this rotation, implement Eq.~\eqref{ex4:drive}
	in a common Hermitian reference space and measure the same fixed
	self-adjoint observable
	$\mathsf A_{\mathrm{ref}}
	=\widetilde{\mathsf H}_{\mathrm{lat}}(0)/\varepsilon_{\mathrm{lat}}$
	after either protocol.
	Writing
	$\langle\mathsf A_{\mathrm{ref}}\rangle_{T,\nu}
	=\langle\psi_\nu(T)|\mathsf A_{\mathrm{ref}}|\psi_\nu(T)\rangle$,
	the adiabatic endpoint expectations are
	\begin{equation}
		\begin{aligned}
			\langle\mathsf A_{\mathrm{ref}}\rangle_{T,\mathrm{lat}}
			&\longrightarrow-1,
			\\
			\langle\mathsf A_{\mathrm{ref}}\rangle_{T,\mathrm b}
			&\longrightarrow
			-\frac{
				m_{\mathrm{lat}}^2+
				J_{\mathrm{lat}}^2\cos(2\zeta_*)
			}{\varepsilon_{\mathrm{lat}}^2}.
		\end{aligned}
		\label{ex4:readout}
	\end{equation}
	The measurement axis is kept fixed.
	For $m_{\mathrm{lat}}/J_{\mathrm{lat}}=1/2$ and $\xi_*=3/5$,
	$\zeta_*=\pi/4$ and the two values are $-1$ and $-1/5$.
	Here $W_{\mathrm b}$ fixes the endpoint rotation relative
	to the common Hermitian measurement axis.
	
	An interferometric comparison additionally resolves the accumulated
	relative phase.
	Define $\vartheta_{\mathrm{lat}}\in(0,\pi)$ by
	$\cos\vartheta_{\mathrm{lat}}
	=m_{\mathrm{lat}}/\varepsilon_{\mathrm{lat}}$
	and
	$\sin\vartheta_{\mathrm{lat}}
	=J_{\mathrm{lat}}/\varepsilon_{\mathrm{lat}}$.
	An instantaneous lower eigenvector of the proper image is
	$U_{\mathrm{lat}}|u_-(\kappa)\rangle$, where
	$|u_-(\kappa)\rangle
	=(-e^{-\mathrm{i}\kappa}\sin(\vartheta_{\mathrm{lat}}/2),
	\cos(\vartheta_{\mathrm{lat}}/2))^{\mathsf T}$.
	Here ${}^{\mathsf T}$ denotes transpose, and we choose
	$|u_0\rangle=|u_-(0)\rangle$.
	Since
	$U_{\mathrm{lat}}^\dagger\dot U_{\mathrm{lat}}
	=-\mathrm{i}f_{\mathrm{lat}}\dot\kappa\sigma_z$,
	its Berry connection is
	\begin{equation}
		\begin{aligned}
			&\mathrm{i}
			\langle U_{\mathrm{lat}}u_-|
			\partial_t(U_{\mathrm{lat}}u_-)\rangle
			\\
			&\quad=
			\left[
			\frac12\left(
			1-\frac{m_{\mathrm{lat}}}{\varepsilon_{\mathrm{lat}}}
			\right)
			-f_{\mathrm{lat}}(\xi)
			\frac{m_{\mathrm{lat}}}{\varepsilon_{\mathrm{lat}}}
			\right]\dot\kappa.
		\end{aligned}
		\label{ex4:berry}
	\end{equation}
	The vertical segments contribute zero.
	Let $\delta\gamma_{\mathrm{lat}}$ denote the geometric phase
	accumulated along $\mathcal C_{\mathrm b}$ relative to
	$\mathcal C_{\mathrm{lat}}$ in this continuous eigenvector convention.
	For equal durations $T$, the common dynamical phase
	$\varepsilon_{\mathrm{lat}}T$ cancels, giving
	\begin{equation}
		\alpha_{\mathrm b}-\alpha_{\mathrm{lat}}
		=\delta\gamma_{\mathrm{lat}}
		=-\zeta_*
		\frac{m_{\mathrm{lat}}}{\varepsilon_{\mathrm{lat}}}
		\quad(\mathrm{mod}\ 2\pi).
		\label{ex4:phase}
	\end{equation}
	The stated continuous eigenvectors fix the phase convention;
	the endpoint alignment below specifies its interferometric readout.
	
	Prepare a control qubit in $(|0\rangle+|1\rangle)/\sqrt2$,
	with the system initially in $|u_0\rangle$.
	Let $\widetilde{\mathsf H}^{\mathrm{drv}}_\nu(t)$ denote
	Eq.~\eqref{ex4:drive} evaluated along $\mathcal C_\nu$,
	where $\nu\in\{\mathrm{lat},\mathrm b\}$.
	Drive the two branches with
	$|0\rangle\langle0|\otimes
	\widetilde{\mathsf H}^{\mathrm{drv}}_{\mathrm{lat}}
	+|1\rangle\langle1|\otimes
	\widetilde{\mathsf H}^{\mathrm{drv}}_{\mathrm b}$.
	After the drives, apply the calibrated physical gate
	\begin{equation}
		\mathsf L_{\mathrm{end}}
		=|0\rangle\langle0|\otimes I
		+|1\rangle\langle1|\otimes W_{\mathrm b}^\dagger.
		\label{ex4:alignment}
	\end{equation}
	This aligns the endpoint states in Eq.~\eqref{ex4:endstates}.
	With its relative gate phase fixed by the same control-qubit
	reference, the control-qubit readouts satisfy
	\begin{equation}
		\begin{aligned}
			\langle\tau_x\rangle
			&\longrightarrow\cos\delta\gamma_{\mathrm{lat}},
			\\
			\langle\tau_y\rangle
			&\longrightarrow\sin\delta\gamma_{\mathrm{lat}}.
		\end{aligned}
		\label{ex4:phase_readout}
	\end{equation}
	Without alignment, the complex coherence contains the extra factor
	$\langle u_0|W_{\mathrm b}|u_0\rangle
	=\allowbreak\cos\zeta_*+
	\allowbreak\mathrm{i}
	(m_{\mathrm{lat}}/\varepsilon_{\mathrm{lat}})\sin\zeta_*$.
	The alignment gate fixes the endpoint identification relative
	to the control-qubit reference.
	Finite-speed transitions can modify the adiabatic readouts.
	
	The chain thus links curvature to endpoint rotations and relative
	phases in proper Hermitian protocols.
	These drives provide a simulation of the specified metric-compatible
	dynamics, with the observable and phase reference fixed by
	the readout apparatus.
	
	\paragraph{Possible experimental protocol for the SSH--RM extension.}
	
	A possible experimental protocol follows from the quasi-Hermitian SSH--RM
	extension defined in Eq.~\eqref{ex4:model}. The system is initialized at
	$(\kappa,\xi)=(0,0)$ in the lower-energy state of
	$\mathsf H_{\mathrm{lat}}(0,0)$, normalized with respect to
	$\eta_{\mathrm{lat}}(0,0)$, and is driven adiabatically along the two paths
	in Eq.~\eqref{ex4:paths}. For each path, the protocol requires the specified
	quasi-Hermitian Hamiltonian, positive metric, and associated metric-compatible
	dynamics, while keeping the initial proper frame fixed. The corresponding
	proper transports are given in Eq.~\eqref{ex4:holonomy}.
	
	At the endpoint, both states are mapped to the common Hermitian reference
	space through their corresponding proper maps, and the fixed self-adjoint
	observable defined in Eq.~\eqref{ex4:readout} is measured. The predicted
	adiabatic endpoint values are $-1$ and
	$-[m_{\mathrm{lat}}^2+J_{\mathrm{lat}}^2\cos(2\zeta_*)]/
	\varepsilon_{\mathrm{lat}}^2$, respectively. For
	$m_{\mathrm{lat}}/J_{\mathrm{lat}}=1/2$ and $\xi_*=3/5$, they become
	$-1$ and $-1/5$. This endpoint contrast provides the primary readout relative
	to the same fixed Hermitian measurement axis.
	
	The relative phase derived above may be investigated with the interferometric
	protocol in Eqs.~\eqref{ex4:alignment} and~\eqref{ex4:phase_readout}, including
	endpoint alignment by $W_{\mathrm b}^{\dagger}$. It is a supplementary
	readout rather than the sole signature of the obstruction, since
	path-dependent Berry phases can also occur in ordinary Hermitian systems.
	
	A direct quasi-Hermitian implementation requires simultaneous engineering
	of the complex hopping amplitudes, complex staggered onsite terms, the positive
	metric $\eta_{\mathrm{lat}}$, and the associated metric-compatible dynamics.
	The feasibility of this combined control is platform dependent. Alternatively,
	direct implementation of the proper Hermitian Hamiltonian in
	Eq.~\eqref{ex4:drive} provides a more direct Hermitian simulation of the
	corresponding positive-metric quasi-Hermitian protocol.

	\section{Conclusions}\label{Sec.6}
	
	Proper similarity transformations for a fixed quasi-Hermitian structure $(H,\eta)$ define a transport problem over parameter space. Their local and global compatibility is encoded in the metric-induced connection.
	
	Nonzero curvature of $\mathcal G_\mu$ obstructs a common proper frame locally. For a flat connection, holonomy around noncontractible loops determines global single-valuedness. Together, flatness and trivial holonomy give the criterion for a global proper frame of $(H,\eta)$.
	
	The same holonomy determines the endpoint orientation of the mapped Hamiltonian. Commuting sewing unitaries leave the Hamiltonian closed, while noncommuting ones rotate its eigenstates relative to a fixed Hermitian measurement axis.
	
	Examples 2 and 3 exhibit these rotations through curvature-dependent circular protocols and winding-dependent transport of a flat connection, respectively. Adiabatic endpoint measurements connect their frame geometry to fixed-reference readouts. Example~4 illustrates the curvature mechanism in a two-sublattice chain. Closed paths with equal winding yield distinct endpoint statistics, and calibrated alignment allows their relative phase to be measured after the common dynamical phase cancels. Examples~3 and~4 therefore suggest possible experimental protocols for the corresponding endpoint readouts.
	
	These comparisons preserve pathwise equivalence while resolving the relations among Hermitian descriptions associated with different control histories.
	
	The metric $\eta$ and the metric-compatible evolution in Eq.~\eqref{5} define the physical framework of these results. The proper Hermitian drives provide a route to simulating these protocols with
	explicit state preparation, control, and readout.
	\begin{acknowledgments}
		H. G. was supported by the Quantum Science and Technology-National
		Science and Technology Major Project (Grant No. 2021ZD0301904) and the National Natural Science
		Foundation of China (Grant No. 12447216). X.-Y. H. was supported by the National Natural Science
		Foundation of China (Grant No. 12405008).
		
		M.-Z.W. and X.-Y.H. contributed equally to this work.
	\end{acknowledgments}
	
	\section*{Data Availability}
	There are no publicly available research data or software supporting this manuscript. Requests for further information or data should be sent to the authors.
	
	\section*{Conflicts of Interest}
	The authors declare no conflict of interest.

	\appendix
	
	\section{PROPERTIES OF $S_p$}\label{appdU}
	
	Starting from the proper condition \eqref{PofS}
	and the decomposition $S_{p} = U \sqrt{\eta}$ with $U$ unitary, a straightforward calculation gives
	\begin{align}\label{eq:formula_052}
		\dot S_{p}^\dagger S_{p}
		&= \sqrt{\eta}\, \dot U^\dagger U \sqrt{\eta} + \dot{\sqrt{\eta}}\, \sqrt{\eta} ,\notag\\
		S_{p}^\dagger \dot S_{p} &= \sqrt{\eta}\, U^\dagger \dot U \sqrt{\eta} + \sqrt{\eta}\, \dot{\sqrt{\eta}} .
	\end{align}
	The proper condition $\dot S_{p}^\dagger S_{p} = S_{p}^\dagger \dot S_{p}$ therefore becomes
	\begin{equation}\label{eq:formula_053}
		\sqrt{\eta}\, \dot U^\dagger U \sqrt{\eta} + \dot{\sqrt{\eta}}\, \sqrt{\eta}
		= \sqrt{\eta}\, U^\dagger \dot U \sqrt{\eta} + \sqrt{\eta}\, \dot{\sqrt{\eta}} .
	\end{equation}
	Since $\sqrt{\eta}$ is invertible, this implies
	\begin{equation}\label{eq:formula_054}
		\dot U^\dagger U + \sqrt{\eta}^{-1} \dot{\sqrt{\eta}} = U^\dagger \dot U + \dot{\sqrt{\eta}}\, \sqrt{\eta}^{-1} .
	\end{equation}
	Moreover, using $\dot U^\dagger U = - U^\dagger \dot U$, one obtains
	\begin{equation}\label{eq:formula_055}
		U^\dagger \dot U = -\frac12 \bigl( \dot{\sqrt{\eta}}\, \sqrt{\eta}^{-1} - \sqrt{\eta}^{-1} \dot{\sqrt{\eta}} \bigr) .
	\end{equation}
	or equivalently
	\begin{equation}\label{A5}
		\dot U = -\frac12 U \bigl( \dot{\sqrt{\eta}}\, \sqrt{\eta}^{-1} - \sqrt{\eta}^{-1} \dot{\sqrt{\eta}} \bigr) .
	\end{equation}
	This is exactly Eq.~\eqref{dU}.

	Any two proper forms for the same $\eta$ along a fixed trajectory differ by a unitary matrix that is constant on that trajectory.
	Let $S_{p,1} = U_1 \sqrt{\eta}$ and $S_{p,2} = U_2 \sqrt{\eta}$ be two proper decompositions. Both $U_1$ and $U_2$ satisfy the differential equation \eqref{A5} (or Eq.~\eqref{dU}). Taking the Hermitian conjugate of Eq.~\eqref{A5} gives $\dot U_i^\dagger = \mathcal{G} U_i^\dagger$ ($i=1,2$) with $\mathcal{G} = \frac12(\dot{\sqrt{\eta}}\sqrt{\eta}^{-1} - \sqrt{\eta}^{-1}\dot{\sqrt{\eta}})$. Then
	\begin{align}\label{eq:formula_056}
		\frac{\mathrm{d}}{\mathrm{d}t}(U_2 U_1^\dagger) =& \dot U_2 U_1^\dagger + U_2 \dot U_1^\dagger\notag\\ =& (-U_2 \mathcal{G}) U_1^\dagger + U_2 (\mathcal{G} U_1^\dagger) = 0 .
	\end{align}
	Hence $U_2U_1^\dagger=C$ is constant along the trajectory, and $S_{p,2}=CS_{p,1}$.

	\section{DETAILS OF THE DIRECTIONAL BERRY-CURVATURE MISMATCH}
	\label{app:directional_mismatch}

	Below we derive the explicit form of the directional
	Berry-curvature mismatch introduced in Sec.~IV\,A.
	From
	$\partial_\nu\eta
	=(\partial_\nu S_p^\dagger)S_p
	+S_p^\dagger\partial_\nu S_p$,
	the transverse properness mismatch can equivalently be written as

	\begin{equation}
		\mathcal M_\nu[S_p]
		=
		\frac{1}{2}
		\left[
		S_p^\dagger\partial_\nu S_p
		-
		(\partial_\nu S_p^\dagger)S_p
		\right].
		\label{eq:app_mismatch_antihermitian_form}
	\end{equation}

	It follows immediately that
	$\mathcal M_\nu^\dagger[S_p]=-\mathcal M_\nu[S_p]$.

	Differentiating the transverse mismatch with respect to $R^\mu$ and
	using the directional proper condition along $\mu$, together with the
	commutativity of mixed partial derivatives, one finds

	\begin{align}
		\partial_\mu\mathcal M_\nu[S_p]
		=
		\partial_\mu S_p^\dagger\partial_\nu S_p-
		\partial_\nu S_p^\dagger\partial_\mu S_p .
		\label{eq:app_transverse_mismatch_derivative}
	\end{align}

	Equation~\eqref{eq:compact_curvature_difference} gives

	\begin{equation}
		\Delta_{n,\mu\nu}
		=
		-\frac{\mi}{2}\partial_\mu
		\langle\Psi_n|
		\mathcal M_\nu[S_p]
		|\Psi_n\rangle .
		\label{eq:app_compact_delta}
	\end{equation}

	Expanding the derivative and using the anti-Hermiticity of
	$\mathcal M_\nu[S_p]$ together with the relation above, we obtain

	\begin{align}
		\Delta_{n,\mu\nu}
		=
		\operatorname{Im}&\Bigl[
		\langle\Psi_n|
		\partial_\mu S_p^\dagger
		\partial_\nu S_p
		|\Psi_n\rangle
		+
		\langle\partial_\mu\Psi_n|
		S_p^\dagger\partial_\nu S_p
		|\Psi_n\rangle
		\nonumber\\
		&-
		\frac{1}{2}
		\langle\partial_\mu\Psi_n|
		\partial_\nu\eta
		|\Psi_n\rangle
		\Bigr].
		\label{eq:app_expanded_delta}
	\end{align}

	Finally, we verify explicitly that the curvature mismatch vanishes when
	the same transformation is proper along both parameter directions.
	If $S_p$ is also proper along $\nu$, then

	\begin{equation}
		(\partial_\nu S_p^\dagger)S_p
		=
		S_p^\dagger\partial_\nu S_p
		=
		\frac{1}{2}\partial_\nu\eta .
		\label{eq:app_transverse_proper}
	\end{equation}

	Differentiating the proper conditions along $\mu$ and $\nu$ and
	antisymmetrizing gives

	\begin{equation}
		\partial_\mu S_p^\dagger\partial_\nu S_p
		=
		\partial_\nu S_p^\dagger\partial_\mu S_p .
		\label{eq:app_cross_derivative_relation}
	\end{equation}

	The right-hand side is the Hermitian conjugate of the left-hand side,
	so $\partial_\mu S_p^\dagger\partial_\nu S_p$ is Hermitian.
	Moreover, the last two terms in
	Eq.~\eqref{eq:app_expanded_delta} cancel by the proper condition along
	$\nu$. Therefore
	$\Delta_{n,\mu\nu}=0$ and
	$\widetilde{\mathcal F}_{n,\mu\nu}
	=\mathcal F_{n,\mu\nu}$, consistently with the result in
	Sec.~IV\,A.

	\section{PROPERTIES OF $\mathcal{K}_\mu$ AND EQUIVALENCE OF THE TWO OBSTRUCTIONS}\label{app2}
	
	In this Appendix, the relations involving a parameter direction $\mu$
	are understood in the pathwise sense along that direction. No common
	multidirectional proper transformation is assumed unless explicitly stated.
	
	From $S_{p} = U \sqrt{\eta}$ and Eq.~\eqref{dU2}, one finds
	\begin{align}\label{B1}
		\partial_\mu S_{p} &= \partial_\mu U \sqrt{\eta} + U \partial_\mu\sqrt{\eta}\notag\\& = -U \mathcal{G}_\mu \sqrt{\eta} + U \partial_\mu\sqrt{\eta} \notag\\
		&= \frac{U}{2} \partial_\mu\sqrt{\eta} + \frac{U}{2} \sqrt{\eta}^{-1}\partial_\mu\sqrt{\eta}\,\sqrt{\eta}
	\end{align}
	With $\mathcal{K}_\mu = \frac12 \eta^{-1} \partial_\mu \eta$, one can verify that
	\begin{align}\label{B2}
		S_{p} \mathcal{K}_\mu %&= U\sqrt{\eta} \cdot \frac12 \eta^{-1} \partial_\mu\eta \notag\\
		%&= \frac{U}{2} \sqrt{\eta}^{-1} \partial_\mu(\sqrt{\eta}\sqrt{\eta}) \\
		&= \frac{U}{2} \sqrt{\eta}^{-1} \bigl( (\partial_\mu\sqrt{\eta})\sqrt{\eta} + \sqrt{\eta}\,\partial_\mu\sqrt{\eta} \bigr)\notag \\
		&= \frac{U}{2} \partial_\mu\sqrt{\eta} + \frac{U}{2} \sqrt{\eta}^{-1}\partial_\mu\sqrt{\eta}\,\sqrt{\eta} .
	\end{align}
	Comparing \eqref{B1} and \eqref{B2}, we obtain $\partial_\mu S_{p} = S_{p} \mathcal{K}_\mu $.
	
	To prove the identity \eqref{22}, we compute
	\begin{align}\label{eq:formula_057}
		\partial_\mu \mathcal{K}_\nu = -\frac12 \eta^{-1} (\partial_\mu\eta) \eta^{-1} \partial_\nu\eta + \frac12 \eta^{-1} \partial_\mu\partial_\nu\eta ,
	\end{align}
	where we have used $\partial_\mu\eta^{-1} = -\eta^{-1}(\partial_\mu\eta)\eta^{-1}$. Similarly,
	\begin{equation}\label{eq:formula_058}
		\partial_\nu \mathcal{K}_\mu = -\frac12 \eta^{-1} (\partial_\nu\eta) \eta^{-1} \partial_\mu\eta + \frac12 \eta^{-1} \partial_\nu\partial_\mu\eta .
	\end{equation}
	Subtracting the two equations and employing $\partial_\mu\partial_\nu\eta = \partial_\nu\partial_\mu\eta$, we obtain
	\begin{equation}\label{eq:formula_059}
		\partial_\mu \mathcal{K}_\nu - \partial_\nu \mathcal{K}_\mu = -\frac12 \eta^{-1} \bigl( (\partial_\mu\eta)\eta^{-1}(\partial_\nu\eta) - (\partial_\nu\eta)\eta^{-1}(\partial_\mu\eta) \bigr) .
	\end{equation}
	The right-hand side of Eq.~\eqref{eq:formula_059} equals
	$-2[\mathcal{K}_\mu,\mathcal{K}_\nu]$, since
	\begin{equation}\label{eq:formula_060}
		[\mathcal{K}_\mu, \mathcal{K}_\nu] = \frac14 \bigl( \eta^{-1}(\partial_\mu\eta) \eta^{-1}(\partial_\nu\eta) - \eta^{-1}(\partial_\nu\eta) \eta^{-1}(\partial_\mu\eta) \bigr) .
	\end{equation}
	This proves Eq.~\eqref{22}.
	
	We now demonstrate the equivalence between the two integrability conditions
	$\mathcal{F}^{\mathcal{K}}_{\mu\nu}=0$ and
	$\mathcal{F}^{\mathcal{G}}_{\mu\nu}=0$ by comparing the infinitesimal
	mismatch between the two possible orders of pathwise continuation.
	Here $[\partial_\mu,\partial_\nu]S_p$ denotes the infinitesimal
	mismatch between the two ordered pathwise continuations,
	$\mu\!\to\!\nu$ and $\nu\!\to\!\mu$, rather than the commutator of
	ordinary mixed partial derivatives of an already existing smooth
	multidirectional section $S_p(\mathbf R)$.
	The same interpretation applies to $[\partial_\mu,\partial_\nu]U$.
	
	First, using the pathwise relations
	$\partial_\mu S_p=S_p\mathcal{K}_\mu$ and
	$\partial_\nu S_p=S_p\mathcal{K}_\nu$, the continuation mismatch is
	\begin{align}\label{eq:formula_061}
		[\partial_\mu,\partial_\nu]S_{p} = \partial_\mu(S_{p}\mathcal{K}_\nu) - \partial_\nu(S_{p}\mathcal{K}_\mu) = S_{p} \mathcal{F}^{\mathcal{K}}_{\mu\nu}.
	\end{align}
	Second, using $S_p=U\sqrt{\eta}$, the smooth single-valuedness of
	$\sqrt{\eta}$, and the pathwise relation
	$\partial_\mu U=-U\mathcal{G}_\mu$, the same continuation mismatch can
	be written as
	\begin{equation}\label{eq:formula_062}
		[\partial_\mu,\partial_\nu]S_{p} = \bigl([\partial_\mu,\partial_\nu]U\bigr)\sqrt{\eta}=-U \mathcal{F}^{\mathcal{G}}_{\mu\nu} \sqrt{\eta}.
	\end{equation}
	The two evaluations lead to
	\begin{equation}\label{eq:formula_063}
		S_{p} \mathcal{F}^{\mathcal{K}}_{\mu\nu} = -U \mathcal{F}^{\mathcal{G}}_{\mu\nu} \sqrt{\eta}.
	\end{equation}
	Inserting $S_{p}=U\sqrt{\eta}$ into Eq.~\eqref{eq:formula_063} and left-multiplying by $U^\dagger$ immediately yields Eq.~\eqref{eq:formula_012}, thereby proving the equivalence of the two operator-level integrability conditions.
	\bibliography{final_references_complete}
\end{document}